\pdfoutput=1  
\def\nk{n_\mathrm{K}}
\def\acap{\\ \nonumber \\}

\def\rfr#1{Equation\,(\ref{#1})}
\def\rfrs#1#2{Equations\,(\ref{#1})--(\ref{#2})}
\def\Rfr#1{Equation\,(\ref{#1})}

\def\derp#1#2{\rp{\partial{#1}}{\partial{#2}}}

\def\virg#1{``#1"}
\def\eqi{\begin{equation}}
\def\eqf{\end{equation}}
\def\rp#1#2{\frac{#1}{#2}}
\def\lb#1{\label{#1}}

\def\ton#1{\left(#1\right)}
\def\qua#1{\left[#1\right]}
\def\grf#1{\left\{#1\right\}}

\RequirePackage[2020-02-02]{latexrelease}
\documentclass{aastex631}
\usepackage{morefloats}
\usepackage[title]{appendix}
\usepackage{textcomp}
\usepackage{booktabs}
\usepackage{multirow}
\usepackage{rotating,tabularx}
\usepackage{float}
\usepackage{enumerate}
\usepackage{rotating}
\usepackage[polutonikogreek,english]{babel}
\usepackage{amsmath,starfont,textgreek,w-greek}
\usepackage[flushleft]{threeparttable}
\usepackage{amsthm}
\usepackage{bigints}
\usepackage{amscd}
\usepackage[mathlines]{lineno}
\usepackage{amssymb,dsfont}
\usepackage{graphicx,epsfig}
\usepackage{natbib}  
\usepackage{txfonts}
\usepackage{xr-hyper}
\usepackage{hyperref}
\usepackage[utf8]{inputenc}
\usepackage{newunicodechar,graphicx}

\usepackage[nointegrals]{wasysym}
\usepackage[caption=false]{subfig}

\allowdisplaybreaks

\makeatletter
\DeclareRobustCommand\ref{%
    \@ifstar\@refstar\T@ref
  }%
  \DeclareRobustCommand\pageref{%
    \@ifstar\@pagerefstar\T@pageref
  }%
 \makeatother

\begin{document}

\title{Will LAGEOS and LARES 2 succeed in accurately measuring frame--dragging?}

\shortauthors{L. Iorio}

\author[0000-0003-4949-2694]{Lorenzo Iorio}
\affiliation{Ministero dell' Istruzione e del Merito. Viale Unit\`{a} di Italia 68, I-70125, Bari (BA),
Italy}

\email{lorenzo.iorio@libero.it}

\begin{abstract}
The current LAGEOS--LARES 2 experiment aims to accurately measure the general relativistic Lense--Thirring effect in the gravitomagnetic field of the spinning Earth generated by the latter's angular momentum $\boldsymbol{J}$. The key quantity to a priori analytically assess the overall systematic uncertainty is the ratio $\mathcal{R}^{J_2}$ of the sum of the classical precessions of the satellites' nodes $\Omega$ induced by the Earth's oblateness $J_2$ to the sum of their post--Newtonian counterparts. \textit{In principle}, if the sum of the inclinations $I$ of both satellites were \textit{exactly} $180^\circ$, the semimajor axes $a$ and the eccentricities $e$ being \textit{identical}, $\mathcal{R}^{J_2}$ would \textit{exactly} vanish. Actually, it is \textit{not} so by a large amount because of the departures of the \textit{real} satellites' orbital configurations from \textcolor{black}{their} \textit{ideal} ones. Thus, $J_2$ impacts not only directly through its own uncertainty, but also \textit{indirectly} through  the errors in all the other physical and orbital parameters entering $\mathcal{R}^{J_2}$. The consequences of this fact are examined in greater details than done so far in the literature. \textcolor{black}{The} Van Patten and Everitt's  proposal in 1976 of looking at the sum of the node precessions of two counter--orbiting spacecraft in (low-altitude) circular polar orbits is revamped rebranding it POLAr RElativity Satellites (POLARES). Regardless  the specific type of satellite and tracking technologies that may be eventually adopted, it \textit{might} be \textit{conceptually} superior to the LAGEOS--LARES 2 one from the point of view of the orbital characteristics since, given the \textit{same} semimajor axes and eccentricities of the existing laser--ranged cousins, its $\mathcal{R}^{J_2}$ is \textit{less} sensitive to the impact of the deviations from its \textit{ideal} orbital configuration.
\end{abstract}


%
%
%
\section{Introduction}
To its first post--Newtonian (1pN) order \citep{2014grav.book.....P}, the General Theory of Relativity (GTR) predicts, among other things, that any mass--energy currents of a source that moves in some way contribute its overall gravitational field with an own, peculiar term encoded in the off--diagonal elements $g_{0i},\,i=1,2,3$ of the spacetime metric tensor $g_{\mu\nu},\,\mu\,\nu=0,1,2,3$. In the case of an isolated rigidly rotating body, such an additional component of its gravitational field, dubbed as gravitomagnetic \citep{Thorne86,1986hmac.book..103T,1988nznf.conf..573T,2001rfg..conf..121M,2001rsgc.book.....R,Mash07}, is proportional to its spin angular momentum $\boldsymbol{J}$. The previous denomination has nothing to do with the magnetic field induced by electric currents, being simply due to the formal resemblance of the linearized  Einstein's field equations in the weak--field and slow--motion limit with those of the Maxwellian electromagnetism. Among the variety of gravitomagnetic phenomena \citep{1986SvPhU..29..215D,2002NCimB.117..743R,2004GReGr..36.2223S,2009SSRv..148...37S}, there is also the so--called Lense--Thirring (LT) effect \citep{1918PhyZ...19..156L,1984GReGr..16..711M} pertaining the orbit a test particle in geodesic motion around its spinning primary. According to recent historical studies \citep{2007GReGr..39.1735P,2008mgm..conf.2456P,Pfister2014}, it would be more appropriate to call it the Einstein--Thirring--Lense effect. Be that as it may, it consists of generally small cumulative changes of the orientation of the orbital plane in space and of the orbit in the orbital plane itself, being its shape and  size left unaltered \citep{2024gpno.book.....I}.

By restricting to our solar system, proposals to measure the LT orbital precessions of natural and artificial bodies revolving about, e.g., the Sun and Jupiter were put forth over the past years; for an overview, see \citep{2011Ap&SS.331..351I}, and references therein. To date, some rare, still inconclusive tests have been performed with Mercury in the solar field \citep{2015IAUGA..2227771P,Pav2024,RussiLT} and with the spacecraft Juno \citep{2017SSRv..213....5B} around Jupiter \citep{2011AGUFM.P41B1620F,2024ApJ...971..145D}.

The situation looks quite different with regards to the Earth in various respects. The idea of using artificial satellites to measure the LT effect in the terrestrial gravitomagnetic field dates back to the late 1950s \citep{Ginz57a,Bogo59,Ginz59}. Actual attempts to detect the gravitomagnetic orbital precessions have been underway since the mid--1990s \citep{1996NCimA.109..575C}  by monitoring the motion of some passive geodetic satellites \citep{2019JGeod..93.2181P} tracked with the Satellite Laser Ranging (SLR) technique \citep{SLR11}, as was first proposed to be done with LAGEOS \citep{1985JGR....90.9217C} by Cugusi and Proverbio \citep{1978A&A....69..321C}. For a comprehensive overview, see, e.g., \citep{2013CEJPh..11..531R}, and references therein. In this respect, among such spacecraft, a prominent role is currently played by LAGEOS, in orbit since 1976, and its cousin LARES 2 \citep{2019JGeod..93.2437P}, launched on July 13, 2022. They are dense metallic spheres entirely covered by retroreflectors \citep{2004CeMDA..88..269L} bouncing back the laser pulses routinely sent to them from several ground--based SLR stations  belonging to the International Laser Ranging Service (ILRS) \citep{2002AdSpR..30..135P}. The shape and composition of these satellites greatly reduce the impact of many non--gravitational disturbing accelerations \citep{Nobilibook87}, thus making them the man--made objects that come closest to the concept of test particles in pure geodesic motion.

To date, the only undisputed test of a gravitomagnetic effect is the one carried out with the Gravity Probe B (GP--B) mission \citep{Varenna74,2001LNP...562...52E} in the circumterrestrial space. It measured the Pugh--Schiff precessions of the spins \citep{Pugh59,Schiff60} of four gyroscopes carried onboard a drag--free spacecraft with a $\simeq 19\%$ \textcolor{black}{accuracy} \citep{2011PhRvL.106v1101E,2015CQGra..32v4001E}.

A major source of systematic bias in all the attempts to measure the LT effect with terrestrial satellites is represented by the disturbances induced by the multipolar expansion of the Newtonian component of the Earth's gravity potential accounting for its departure from spherical symmetry \citep{1967phge.book.....H,Torge01}. Indeed, the resulting classical orbital shifts \citep{2003CeMDA..86..277I,2005som..book.....C} have the \textit{same} temporal behaviour of the LT ones, \textcolor{black}{along with} the fact that they exhibit \textit{much larger} nominal magnitudes. Among them, the largest by far are those due to the first even zonal harmonic coefficient $J_2$ of degree $\ell = 2$ and order $m=0$  of the geopotential accounting for the Earth's oblateness.

In order to gain useful insight into the relative sizes of all such competing features of motion, it is convenient to reason in terms of the standard Keplerian orbital elements \citep{2011rcms.book.....K}. In some Earth--centered \textcolor{black}{asymptotic} inertial (ECI) reference frame $\mathcal{K}$ an axis of which is \textit{exactly} aligned with the terrestrial angular momentum, the longitude of the ascending node $\Omega$ of the test particle is displaced by both the gravitomagnetic and the classical quadrupolar gravitational fields according to \citep{1918PhyZ...19..156L,2005som..book.....C}
\begin{align}
\dot\Omega_\mathrm{LT} \lb{OdotLT}& = \rp{2 GJ}{c^2 a^3\ton{1 - e^2}^{3/2}}, \acap
\dot\Omega_{J_2} \lb{OdotJ2}& = -\rp{3}{2}\nk\ton{\rp{R}{p}}^2\cos I,
\end{align}
where $c$ is the speed of light in vacuum, $G$ is the Newtonian constant of gravitation, $a$ is the semimajor axis, $e$ is the eccentricity, $p:=a\ton{1-e^2}$ is the semilatus rectum,  $I$ is, in this case, the inclination of the orbital plane to the equatorial plane of the central body whose mass and equatorial radius are $M$ and $R$, respectively. Furthermore, $\upmu:=GM$ is its standard gravitational parameter, and $\nk:=\sqrt{\upmu/a^3}$ is the Keplerian mean motion of the test particle.
The longitude of the ascending node $\Omega$ is an angle reckoned in the adopted reference plane $\Pi$ of $\mathcal{K}$ from the $x$ direction to the unit vector
\eqi
\boldsymbol{\hat{l}}=\grf{\cos\Omega,\sin\Omega,0}\lb{elle}
\eqf
of the line of nodes, which is the intersection  of the orbital plane with $\Pi$ itself. The versor $\boldsymbol{\hat{l}}$ is directed toward the ascending node $\ascnode$, which is the point where the orbiter crosses $\Pi$ from below.
In principle, the argument of pericentre $\omega$, which is an angle counted in the orbital plane from $\boldsymbol{\hat{l}}$ to the line of apsides oriented toward the point of closest approach, also undergoes, among other things, a secular LT precession \citep{1918PhyZ...19..156L}. Nonetheless, it has been a long time since, after some initial attempts \citep{1996NCimA.109..575C,1997CQGra..14.2701C,1998Sci...279.2100C} involving also the perigee of LAGEOS 2 \citep{1989AcAau..19..521I}, it was decided not to use such an orbital element anymore. Indeed, it is heavily perturbed by a host of non--gravitational accelerations \citep{2001P&SS...49..447L,2002P&SS...50.1067L,2004P&SS...52..699L}.

An remarkable feature of \rfrs{OdotLT}{OdotJ2} is that, if on the one hand, the LT node precession does \textit{not} depend on $I$ at all, on the other hand, the classical rate gains on \textit{opposite} sign if $I$ is switched by $180^\circ$. Thus, if there were two satellites the sum of whose inclinations is \textit{exactly} $180^\circ$, all the other orbital parameters being \textit{identical}, the \textit{sum} of their nodes would allow, \textit{in principle}, to add up the LT rates while the nominally much larger competing precessions due to $J_2$ would \textit{exactly} cancel out. The same property would automatically extend also to \textit{all} the other disturbing Newtonian rates induced the \textit{even zonal} harmonics $J_\ell,\,\ell=4,6,\ldots$ of higher degree since it turns out that they are \textit{all} proportional to \rfr{OdotJ2} through certain functions of $I$ which are left unaffected by the replacement $I\rightarrow 180^\circ - I$ \citep{2003CeMDA..86..277I}.

\begin{figure}
\centering
\includegraphics[width=\columnwidth]{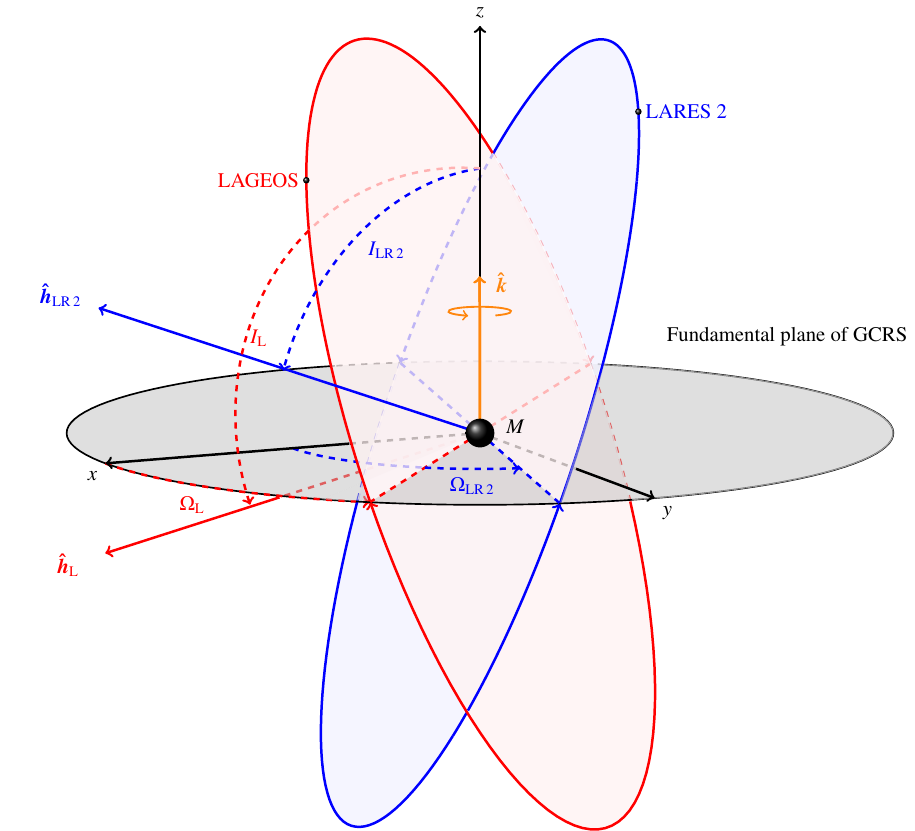}
\caption{Orbital configurations of LAGEOS (in red, L) and LARES 2 (in blue, LR 2) at the epoch $T_1$ of the launch of LARES 2 with respect to the Geocentric Celestial Reference System (GCRS) whose fundamental plane is shown shaded in grey. The Earth's spin axis $\boldsymbol{\hat{k}}$, in orange, is depicted according to \rfr{kT1}. The ratio of the semimajor axis $a$ of LARES 2 to that of LAGEOS,  to which an arbitrary reference value has been assigned just for illustrative purpose, is as in Table \ref{Tab:1} from which the  eccentricities $e$ and inclinations $I$ are retrieved as well. The longitudes of the ascending node $\Omega$ at $T_1$ are as in \rfrs{OL}{OLR}. The unit vectors $\boldsymbol{\hat{h}}$, given by \rfr{acca}, are directed towards the orbital angular momenta of the satellites. The sizes of the Earth and of the satellites' orbits are \textit{not} in scale.
}\label{fig_L_LR2}
\end{figure}

This is just the line followed by Ciufolini and coworkers who, in a series of recent papers \citep{2023EPJC...83...87C,Ciufoepjc24}, claimed to be able to perform in the near future a LT test accurate to $\simeq 0.2\%$ with LAGEOS and LARES 2 whose inclinations are $I^\mathrm{L}\simeq 110^\circ$ and $I^\mathrm{LR2}\simeq 70^\circ$, while their semimajor axes and eccentricities are \textit{almost} identical, as per Table \ref{Tab:1}.
\begin{table}[h]
\caption{Relevant orbital parameters of LAGEOS and LARES 2 \cite[Tab.\,1]{2023EPJC...83...87C}. They are mean values over 127 days.
}\lb{Tab:1}
\begin{center}
\begin{tabular}{|l|l|l|l|}
  \hline
& Semimajor axis $a$ (km)  & Eccentricity $e$ & Inclination $I$ ($^\circ$) \\
\hline
LAGEOS  & $12\,270.020705$ & $0.00403$ & $109.8469$\\
LARES 2 & $12\,266.1359395$ & $0.00027$ & $70.1615$\\
\hline
\end{tabular}
\end{center}
\end{table}
Figure \ref{fig_L_LR2} shows the orbital geometry of LAGEOS and LARES 2.
Relying upon \citep{1986PhRvL..56..278C}, Ciufolini and coworkers look at the \textit{sum} of the node precessions of LAGEOS and LARES 2 in order to extract a sufficiently clean LT signal. In this respect, the key quantity to \textit{analytically} assess the systematic bias due to $J_2$ on the LT signal is the ratio
\eqi
\mathcal{R}^{J_2}:= \rp{\dot\Omega_{J_2}^\mathrm{L} + \dot\Omega_{J_2}^\mathrm{LR2} }{\dot\Omega_\mathrm{LT}^\mathrm{L} + \dot\Omega_\mathrm{LT}^\mathrm{LR2}}\lb{RJ2}
\eqf
of the sum of the classical precessions of LAGEOS \textcolor{black}{(L)} and LARES 2 \textcolor{black}{(LR 2)} to the sum of the LT precessions of the same satellites. In an \textit{ideal} case, characterized by \textit{identical} semimajor axes and eccentricities and by \textit{exactly} supplementary inclinations, $\mathcal{R}^{J_2}$ would  vanish. In fact, as pointed out in \citep{2023Univ....9..211I}, even its \textit{nominal} value is \textit{by far not} zero because of the departures of the \textit{actual} satellites' orbital configurations from the previously mentioned \textit{ideal} scenario. This important fact opens the door to  potentially relevant systematic errors  induced by the uncertainties in \textit{all} the relevant physical and orbital parameters entering the analytical expression of $\mathcal{R}^{J_2}$. In other words, the Earth's oblateness has an impact not only \textit{directly} through its own uncertainty, but also \textit{indirectly} through the errors in the other parameters leaking into the largely nonvanishing expression of \rfr{RJ2}.

Aim of the present work is to examine such an important issue in greater detail than has been done so far in the literature \citep{2023Univ....9..211I,Ciufoepjc24}, and to find possible alternatives by reexamining an earlier proposal put forth in 1976 by Van Patten and Everitt \citep{1976CeMec..13..429V,1976PhRvL..36..629V}.

The paper is organized as follows. Section\,\ref{preceq} treats the consequences of the precession of the Earth's spin axis on $\mathcal{R}^{J_2}$. \textcolor{black}{In particular, Section\,\ref{errGJ} deals with the impact of the uncertainties in those parameters which act as scaling factors on the recalculated expression of the ratio of the summed  classical to relativistic node precessions: $G$ (Section\,\ref{errG}) and $J$ (Section\,\ref{errJ}). The effect of the secular variation of $J_2$ is treated in Section\,\ref{errJ2dot}. A quick assessment of the bias due to $J_4$ is the subject of Section\,\ref{errJ4}}. The scenario envisaged by Van Patten and Everitt is revisited in Section\,\ref{vanPa}. In it, the features of counter--orbiting test particles along identical orbits, however inclined, are reviewed in Section\,\ref{counter} showing its equivalence with the LAGEOS--LARES 2 system, while the case of polar orbits is examined in Section\,\ref{polares}. In Section\,\ref{corbex}, the criticisms by Ciufolini \textit{et al.} \citep{Ciufoepjc24} are addressed. In particular, the issue of the correct error propagation  is the subject of Section\,\ref{errpro}, while Section\,\ref{bsciard} is dedicated to the impact of the uncertainties in the orbital elements: the semimajor axis (Section\,\ref{sema}) and the inclination (Section\,\ref{inclis}).
Section\,\ref{fine} summarizes the findings and offers conclusions.
\section{The direct and indirect consequences of the precession of the Earth's axis on $\mathcal{R}^{J_2}$}\lb{preceq}
The ECI used in satellite geodesy is the Geocentric Celestial Reference System (GCRS) \citep{iers10}. It is essentially characterized, among other things, by the Earth's Mean Equator and Mean Equinox (MEME) at 12:00 Terrestrial Time on 1 January 2000 (J2000.0), being also dubbed as J2000 system \citep{2004sod..book.....T}. Its $x$ axis is aligned with the mean vernal equinox. Its $z$ axis is aligned with the Earth's rotation axis (or equivalently, the celestial North Pole) as it was at \textit{that} time. The $y$ axis is rotated by $90^\circ$ East about the celestial equator \citep{2004sod..book.....T}. More precisely, as per the Recommendation 2 of the IAU 2006 Resolution B.2 by the International Astronomical Union (IAU) \citep{iers10}, the \textit{orientation} of GCRS coincides \textit{by default} with that of the International Celestial Reference System (ICRS). The principal plane $\Pi$ of the latter and its origin are chosen to be as \textit{close as possible} to the Earth's mean equator and equinox at J2000.0; there is a fixed offset, known as frame bias, of about 23 milliarcseconds (mas) between the two systems  \citep{2006A&A...450..855C}. See Section \ref{inclis} for how this impacts the realistic accuracy in knowing the satellites' inclinations.

LARES 2 was launched $21.53$ years \textit{after} the reference epoch J2000.0. Thus, data analyses aimed to detect the LT effect with LAGEOS and LARES 2 will be necessarily carried out starting at least from \textit{that} date onwards. Furthermore, they can generally last for decades.
Actually, it \textit{does matter} since, in the meantime, the Earth's spin axis $\boldsymbol{\hat{k}}$ \textit{has changed} mainly due to several physical processes, the most relevant of which is the precession of the equinoxes having  a period of approximately 26\,000 years \citep{PrecNut}. In fact, the LT and classical orbital precessions \textit{generally} depend on the orientation of $\boldsymbol{\hat{k}}$ with respect to the inertial frame adopted. They are equal to \citep{2024gpno.book.....I}
\begin{align}
\dot\Omega_\mathrm{LT} \lb{OLT}& = \rp{2GJ\csc I}{c^2 a^3\ton{1 - e^2}^{3/2}}\boldsymbol{\hat{k}}\boldsymbol\cdot\boldsymbol{\hat{m}},\acap
\dot\Omega_{J_2} \lb{OJ2}& = -\rp{3}{2}\nk\ton{\rp{R}{p}}^2\csc I\ton{\boldsymbol{\hat{k}}\boldsymbol\cdot\boldsymbol{\hat{m}}}\ton{\boldsymbol{\hat{k}}\boldsymbol\cdot\boldsymbol{\hat{h}}},
\end{align}
where
\begin{align}
\boldsymbol{\hat{m}} \lb{emme}&  =\grf{-\cos I \sin\Omega, \cos I \cos\Omega, \sin I},\acap
\boldsymbol{\hat{h}} \lb{acca}&  =\grf{\sin I \sin\Omega, -\sin I \cos\Omega, \cos I}.
\end{align}
The unit vector $\boldsymbol{\hat{h}}$ is directed along the satellite's orbital angular momentum, while $\boldsymbol{\hat{m}}$ lies in the orbital plane in such a way that $\boldsymbol{\hat{l}}\boldsymbol\times\boldsymbol{\hat{m}} = \boldsymbol{\hat{h}}$  holds.
It should be recalled that \rfrs{OdotLT}{OdotJ2} hold \textit{only} at \textit{J2000.0} when $\boldsymbol{\hat{k}}_\mathrm{J2000.0}=\grf{0,0,1}$. At this point, it may be the case to recall that, for an arbitrary orientation of $\boldsymbol{\hat{k}}$ in space, also the inclination $I$ undergo long--term shifts given by
\begin{align}
\dot I_\mathrm{LT} \lb{ILT}& = \rp{2GJ}{c^2 a^3\ton{1 - e^2}^{3/2}}\boldsymbol{\hat{k}}\boldsymbol\cdot\boldsymbol{\hat{l}},\acap
\dot I_{J_2} \lb{IJ2}& = -\rp{3}{2}\nk\ton{\rp{R}{p}}^2\ton{\boldsymbol{\hat{k}}\boldsymbol\cdot\boldsymbol{\hat{l}}}\ton{\boldsymbol{\hat{k}}\boldsymbol\cdot\boldsymbol{\hat{h}}},
\end{align}

Thus, one may wonder to what extent the expression of $\mathcal{R}^{J_2}$ used by Iorio \citep{2023Univ....9..211I} and Ciufolini \textit{et al.} \citep{Ciufoepjc24}, \textit{calculated with \rfrs{OdotLT}{OdotJ2}}, is adequate.
The answer is negative, as it will be shown in the following.

The first step is using \textit{\rfrs{OLT}{OJ2}} to calculate an expression for $\mathcal{R}^{J_2}$ valid for the \textit{epoch of the launch of LARES 2}, denoted in the following as $T_1$. To this aim, one has first to obtain the Earth's spin axis $\boldsymbol{\hat{k}}_{T_1}$ just at $T_1$. By taking into account only the effect of the precession for simplicity, this task can be accomplished by means of the standard formulas providing the orientation of the mean equator and equinox of the generic epoch $T$ with respect to the MEME \cite[p.\,176]{2000Monte}. It turns out
\eqi
{\boldsymbol{\hat{k}}}_{T_1} = \grf{-0.00209215,-5.04\times 10^{-6},0.99999781}.\lb{kT1}
\eqf
By parameterizing the Earth's spin axis in terms of the right ascension (R.A.) $\alpha$ and declination (decl.) $\delta$ of the Earth's North Pole of rotation as
\eqi
\boldsymbol{\hat{k}} = \grf{\cos\alpha\cos\delta,\sin\alpha\cos\delta,\sin\delta},\lb{kappa}
\eqf
\rfr{kT1} corresponds to
\begin{align}
\alpha_{T_1} \lb{RA}& = 0.13815807^\circ, \acap
\delta_{T_1} \lb{DEC}& = 89.88012829^\circ.
\end{align}

The next step is obtaining the values of the longitudes of the ascending node of LAGEOS and LARES 2 at $T_1$, not provided in \cite[Tab.\,1]{2023EPJC...83...87C}. This can be approximately done by retrieving their values at any epoch by means of, say, the WEB interface provided at \url{https://www.n2yo.com/},
and propagating them backward in time to $T_1$ by means of \rfr{OdotJ2}. Thus, one finally has
\begin{align}
\Omega^\mathrm{L} \lb{OL}& \simeq 49.55^\circ, \acap
\Omega^\mathrm{LR 2} \lb{OLR}& \simeq 76.15^\circ.
\end{align}

By calculating \rfrs{OLT}{OJ2} with \rfr{kT1}, \rfrs{OL}{OLR} and the values of the other orbital parameters listed in Table \ref{Tab:1}, one obtains for \rfr{RJ2}
\eqi
\left.\mathcal{R}^{J_2}\right|_{T_1} \simeq 59\,161.9.\lb{morto}
\eqf
The figure of \rfr{morto} is even about $\mathit{12}$ \textit{times larger} than that calculated in \citep{2023Univ....9..211I} \textit{by means of \rfrs{OdotLT}{OdotJ2}}.
A \textit{nominal} value of $\mathcal{R}^{J_2}$  as large as that provided by \rfr{morto} is potentially a serious issue. Indeed, even relatively small errors in some of the physical and orbital parameters entering \rfr{RJ2} may propagate  inducing too large a systematic bias to meet the accuracy goal stated by Ciufolini \textit{et al.} \citep{2023EPJC...83...87C,Ciufoepjc24}. In this respect, the results by Iorio \citep{2023Univ....9..211I} about the consequences of the uncertainties in $a,e$ and $I$ retain their overall validity, even if a reassessment of the latter ones may be needed; see Sections \ref{sema} to \ref{inclis}.

\begin{figure}[ht!]
\centering
\begin{tabular}{c}
\includegraphics[width = 15 cm]{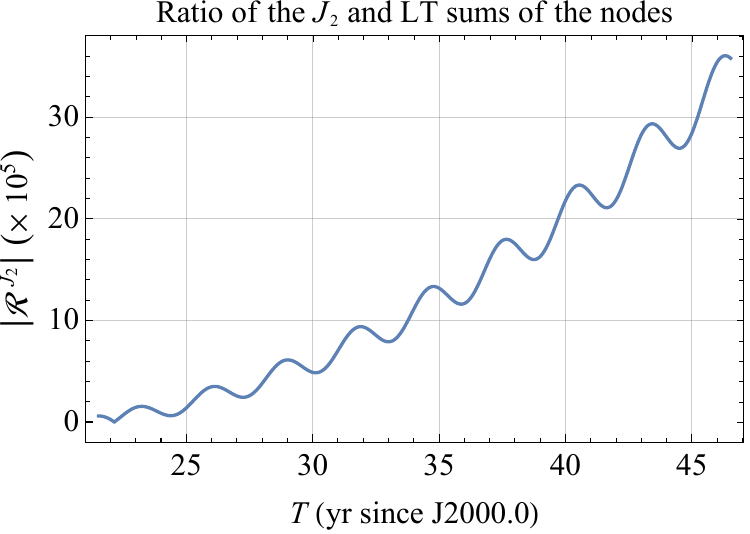}\\
\end{tabular}
\caption{
Plot of the absolute value of \rfr{RJ2}, calculated with the general expressions for the LT and classical node rates of \rfrs{OLT}{OJ2}, as a function of the epoch $T$ over a hypothetical data analysis time span 25 yr long starting from the LARES 2 launch date. The precession of the Earth's spin axis $\boldsymbol{\hat{k}}$, worked out as explained in \cite[pp.\,176-177]{2000Monte}, was included along with the temporal displacements of $I$ and $\Omega$ as per \rfr{IJ2} and \rfr{OJ2}. Also the secular variation of the Earth's oblateness was implemented as per \rfr{J2}.
}\label{Fig:1}
\end{figure}

Furthermore, it must be remarked that, actually, $\mathcal{R}^{J_2}$ is time--dependent due, among other things, to the slow precessional motion of $\boldsymbol{\hat{k}}$. The precession transformation  between arbitrary epochs can be worked out as detailed in \cite[pp.\,176-177]{2000Monte}. By applying the resulting calculation to the Earth's spin axis between $T_1$ and a generic epoch $T_2>T_1$ allows to plot the \textit{nominal} value of $\mathcal{R}^{J_2}$, calculated with \rfrs{OLT}{OJ2}, against $T_2$ assumed as independent variable; see Figure \ref{Fig:1}.
\subsection{\textcolor{black}{The impact of some physical parameters acting as scaling factors}}\lb{errGJ}
\textcolor{black}{Here, the impact of the Newtonian constant of gravitation $G$ and the Earth's angular momentum $J$ is evaluated separately.}

\textcolor{black}{Since both act as scaling factors for the LT node precessions, they could be removed, at least in principle, by using a suitable linear combination of the nodes of LAGEOS, LARES 2 and some other probes like LAGEOS 2. Such an approach, proposed for the first time by Shapiro \citep{1990grg..conf..313S} in a different scenario, was extended years later also to some SLR targets of the LAGEOS family by Ciufolini \citep{1996NCimA.109.1709C}. In particular, the possibility of linearly combining the nodes of LAGEOS, LAGEOS 2 and a satellite which was intended to have the same orbital parameters of the current LARES 2, was explicitly considered by Iorio \textit{et al.} \citep{2002CQGra..19.4311I}. In fact, Ciufolini \textit{et al.} \citep{2023EPJC...83...87C,Ciufoepjc24} have so far only considered the sum of the nodes of LAGEOS and LARES 2, giving no signs of wanting to consider any other alternative. Furthermore, also the aforementioned approach would not be free from drawbacks. Indeed, since the coefficients weighing the residuals of the orbital elements entering the linear combinations are theoretically calculated from some mean values of the satellites' orbital parameters, they would be affect by the uncertainties in the latter ones, thus not allowing for a perfect cancellation of the scaling parameters whose bias one wish to remove \citep{2011GReGr..43.1697I}.}
\subsubsection{The impact of the uncertainty in the Newtonian constant of gravitation}\lb{errG}
Let the impact of the uncertainty in $G$, overlooked by either Iorio \citep{2023Univ....9..211I} and Ciufolini \textit{et al.} \citep{Ciufoepjc24}, be considered.

On the one hand, $G$ enters \rfr{OJ2} only through $\upmu$, which is one of the parameters  that are accurately estimated in data reductions of satellites' SLR observations. On the other hand, the product $GJ$ entering \rfr{OLT} does  $\textit{not}$ (yet?) fall within such a category; thus, the impact of our imperfect knowledge of  $G$ on \rfr{RJ2} must be evaluated \textit{separately}.

According to the 2022 CODATA recommended values, provided  by the National Institute of Standards and Technology (NIST)  and retrievable on the  at \url{https://physics.nist.gov/cgi-bin/cuu/Value?bg}, its relative uncertainty is currently
\eqi
\rp{\sigma_G}{G} = 2.2\times 10^{-5}.\lb{erroG}
\eqf
By scaling \rfr{morto} by \rfr{erroG}, an error in $\mathcal{R}^{J_2}$ as large as
\eqi
\sigma_{\mathcal{R}^{J_2}} \leq \left|\derp{\mathcal{R}^{J_2}}{G}\right|^{T_1}_{\upmu=\mathrm{const}}\sigma_G = \left|\mathcal{R}^{J_2}\right|_{T_1}\rp{\sigma_G}{G} = 1.3,
\eqf
corresponding to a $\simeq 130\%$ systematic bias due to $J_2$ in the expected LT signature, occurs.

Interestingly, even with the J2000.0 expression of $\mathcal{R}^{J_2}$, calculated with \textit{\rfrs{OdotLT}{OdotJ2}} and used in \citep{2023Univ....9..211I,Ciufoepjc24}, \rfr{erroG} would still yield a $\simeq 11\%$ oblateness--driven systematic uncertainty in the sum of the LT node precessions.
\subsubsection{The impact of the uncertainty in the Earth's angular momentum}\lb{errJ}
By citing relevant works, Ciufolini \textit{et al.} \cite[Tab.\,2]{Ciufoepjc24} \textit{convincingly} demonstrated that the realistic relative error in the Earth's angular momentum is actually \textit{smaller} than guessed by Iorio \cite[Sect.\,4]{2023Univ....9..211I}, being of the order of
\eqi
\rp{\sigma_J}{J}\simeq 10^{-6}.\lb{sigmaJ}
\eqf
If, on the one hand, \rfr{sigmaJ} is small enough not to pose problems to the J2000.0 expression of $\mathcal{R}^{J_2}$ used in \citep{2023Univ....9..211I,Ciufoepjc24}, on the other hand it is not entirely so for \rfr{morto}. Indeed, by rescaling it by \rfr{sigmaJ} one has
\eqi
\sigma_{\mathcal{R}^{J_2}}\leq \left|\derp{\mathcal{R}^{J_2}}{J}\right|\sigma_J = \left|\mathcal{R}^{J_2}\right|_{T_1}\rp{\sigma_J}{J}  = 0.059,
\eqf
corresponding to a $\simeq 6\%$ bias.
\subsection{The impact of the uncertainty in the secular variation of the Earth's oblateness}\lb{errJ2dot}
The first even zonal harmonic $J_2$ of the geopotential is brought in the numerator of \rfr{RJ2} by \rfr{OJ2} as a multiplicative parameter  common to  the node rates of both LAGEOS and LARES 2.
%
%
%

Indeed,  a variety of physical processes induce  certain time--dependent variations of the Earth's oblateness which can be expressed as
\eqi
J_2\ton{T} = \overline{J}_2 + \dot J_2\ton{\rp{T - T_\mathrm{rf}}{P}} + J_2^\mathrm{c}\cos\qua{\rp{2\uppi}{P}\ton{T - T_\mathrm{rf}}} +  J_2^\mathrm{s}\sin\qua{\rp{2\uppi}{P}\ton{T - T_\mathrm{rf}}},\lb{J2}
\eqf
where $\overline{J}_2$ is meant here as the coefficient of degree $\ell=2$ and order $m=0$ of the unconstrained static field, $\dot J_2$ is the amplitude of the linear trend, in yr$^{-1}$, $J_2^\mathrm{c}$ and $J_2^\mathrm{s}$ are the amplitudes of the harmonic annual variations, $T$ is an arbitrary epoch,  $T_\mathrm{rf}$ is some reference epoch depending on the Earth's gravity field solution adopted, $P=365.25\,\mathrm{d}$ is the duration of the year, in days. As an example, for the model ITSG-Grace2018, obtained from 162 months of GRACE\footnote{\textcolor{black}{Values of the low-degree even zonals from Earth's gravity solutions obtained only from SLR satellites like, e.g., IGG$\_$UPWr$\_$SLR retrievable at \url{https://doi.org/10.1016/j.rse.2024.113994}, would not be suitable since they would be a priori imprinted just by the LT effect, not explicitly solved-for in them.}}. data collected from April 2002 to June 2017 and retrievable at \url{http://doi.org/10.5880/ICGEM.2018.003}, the reference epoch $T_\mathrm{rf}$ is June 1, 2010.

Should only the static component of $J_2$ affect \rfr{RJ2}, as seemingly assumed by \citep{Ciufoepjc24}, no problems would arise, at least at first sight. Indeed, the present--day level of the \textit{formal} relative uncertainty in $\overline{J}_2$ is at the
\eqi
\rp{\sigma_{\overline{J}_2}}{\overline{J}_2}\simeq 10^{-9}\lb{itsg}
\eqf
level, as for  ITSG-Grace2018. However, Ciufolini \textit{et al.} \citep{Ciufoepjc24} conservatively considered the \textit{calibrated} relative uncertainty in the static part of $J_2$ released by the model GGM05S \citep{Riesetal16}
\eqi
\rp{\sigma_{\overline{J}_2}}{\overline{J}_2}\simeq 2.4\times 10^{-7}.\lb{ggm05}
\eqf
By rescaling \rfr{morto} by \rfr{ggm05}, one gets a systematic bias on the summed LT node precessions of nearly $1.4\%$, which is nearly one order of magnitude larger than the overall accuracy level claimed by Ciufolini \textit{et al.} \citep{Ciufoepjc24}.

In fact, also the mismodeling in the \textit{other} components of \rfr{J2} has, in principle, to be taken into account. Limiting just to the secular trend of $J_2$, it yields a correction $\Delta J_2$ to the static value of the Earth's oblateness which, at the epoch $T_1$ of the LARES 2 launch, is
\eqi
\Delta J_2\ton{T_1}  \simeq  6.3\times 10^{-10}\lb{DJ2}.
\eqf
The relative uncertainty in \rfr{DJ2}, retrievable from the published error in $\dot J_2$ according to, e.g., ITSG-Grace2018, turns out to be as large as
\eqi
\left.\rp{\sigma_{\Delta J_2}}{\Delta J_2}\right|_{T_1} = \left|\rp{\sigma_{\dot J_2}}{\dot J_2}\right|\simeq 1.4\times  10^{-2}.\lb{kill}
\eqf
It corresponds to an error in $\mathcal{R}^{J_2}$, calculated by replacing $J_2\rightarrow \Delta J_2\ton{T_1}$ in \rfr{OJ2} entering \rfr{RJ2}, of the order of
\eqi
\sigma_{\mathcal{R}^{J_2}}\leq \left|\derp{\mathcal{R}^{J_2}}{\Delta J_2}\right|_{T_1}\sigma_{\Delta J_2}  \simeq 5\times 10^{-4}.
\eqf
However, due to the non--trivial overall time dependence of $\mathcal{R}^{J_2}$ introduced by the precessional motion of $\boldsymbol{\hat{k}}$ and by the temporal evolution of $I$ and $\Omega$, it would be advisable to fully account also for $\dot J_2$.
Its \textit{nominal} impact  on $\mathcal{R}^{J_2}$ was accounted for in producing Figure \ref{Fig:1}.
%
%
%
\section{\textcolor{black}{The impact of the other Earth's gravity field multipoles}}\lb{errJ4}
\textcolor{black}{In principle, also the other even zonal harmonics $J_\ell,\,\ell=4,6,\ldots$ of higher degree should be taken into account.}

\textcolor{black}{A quick evaluation about the nominal impact of $J_4\simeq 10^{-7}$ of the sum of the nodes can be easily inferred  simply by rescaling \rfr{RJ2} by}
\eqi
\textcolor{black}{\rp{J_4}{J_2}\ton{\rp{R}{a}}^2 \simeq 4\times 10^{-4}.}\lb{factJ4}
\eqf
\textcolor{black}{Thus, from \rfr{morto} it can be guessed}
\eqi
\textcolor{black}{\mathcal{R}^{J_4}\simeq 23.9.}\lb{RJ4}
\eqf
\textcolor{black}{
An application of the scaling factor of \rfr{factJ4} to the time series of Figure\,\ref{Fig:1} clearly shows that a $\simeq 10^3$ bias due to the imperfectly cancelled  node precessions due to $J_4$ would indirectly occur over the years because of the Earth's spin axis precession.
}
\section{Revisiting the van Patten--Everitt proposal for two counter-orbiting polar satellites}\lb{vanPa}
In 1976, Van Patten and Everitt \citep{1976CeMec..13..429V,1976PhRvL..36..629V} suggested to measure the LT effect by using a pair of low--altitude counter--orbiting drag--free Earth's satellites A and B in circular polar motion. In addition to tracking data from existing ground stations, satellite--to--satellite Doppler ranging data should have been taken near the poles.

Apart from \textit{obvious} differences in terms of \textit{cost} and \textit{technologies} to be employed, such a proposal is \textit{conceptually} equivalent to that put forth by  Ciufolini \citep{1986PhRvL..56..278C} \textit{ten} years later (see Section \ref{counter}) and, within certain limits, \textit{even better} from the point of view of the overall accuracy because of the unavoidable departures from the \textit{ideal} orbital configuration (see Section \ref{polares}).
\subsection{Counter--revolving satellites along identical, arbitrarily inclined orbits}\lb{counter}
In fact, the orbits of A and B may not necessarily pass through $\boldsymbol{J}$  in order to yield a \textit{conceptually} equivalent scenario to that envisaged by Ciufolini \citep{1986PhRvL..56..278C}. This is proven as follows.

By assuming \textit{ideally} identical semimajor axes and eccentricities, a satellite B is counter--revolving with respect to another satellite A if the conditions
\begin{align}
I_\mathrm{B} \lb{coI} & = 180^\circ -  I_\mathrm{A}, \acap
\Omega_\mathrm{B} \lb{coO} & = \Omega_\mathrm{A} + 180^\circ
\end{align}
\textit{exactly} hold.
Indeed, from \rfr{elle} and \rfrs{emme}{acca}, along with the trigonometric identities
\begin{align}
\sin\ton{180^\circ - \beta} \lb{sinpx}& = \sin\beta,\acap
\cos\ton{180^\circ - \beta} \lb{cospx}& = -\cos\beta,\acap
\sin\ton{180^\circ + \beta} \lb{sinxp}& = -\sin\beta,\acap
\cos\ton{180^\circ + \beta} \lb{cosxp}& = -\cos\beta,
\end{align}
 it turns out that
\begin{align}
{\boldsymbol{\hat{l}}}_\mathrm{B} \lb{elleAB}& = -{\boldsymbol{\hat{l}}}_\mathrm{A},\acap
{\boldsymbol{\hat{m}}}_\mathrm{B} \lb{emmeAB}& = {\boldsymbol{\hat{m}}}_\mathrm{A},\acap
{\boldsymbol{\hat{h}}}_\mathrm{B} \lb{accaAB}& = -{\boldsymbol{\hat{h}}}_\mathrm{A}.
\end{align}
\Rfr{accaAB} implies that the sense of motion of B along its orbit is just \textit{opposite} to that of A, while \rfrs{elleAB}{emmeAB}, together with \rfrs{coI}{coO}, guarantee that the orbital planes of A and B \textit{coincide}. Finally, \rfrs{elleAB}{accaAB} yield
\eqi
{\boldsymbol{\hat{l}}}_\mathrm{B}\boldsymbol\times{\boldsymbol{\hat{m}}}_\mathrm{B} = {\boldsymbol{\hat{m}}}_\mathrm{A}\boldsymbol\times{\boldsymbol{\hat{l}}}_\mathrm{A} = -{\boldsymbol{\hat{h}}}_\mathrm{A} = {\boldsymbol{\hat{h}}}_\mathrm{B}.
\eqf
It should be noted that, so far, \textit{no} assumptions on the mutual orientation of $\boldsymbol{\hat{k}}$ and $\boldsymbol{\hat{h}}_\mathrm{A/B}$, were made at all; in other words, the condition of passage through the primary's poles was \textit{not} adopted.
From \rfrs{OLT}{OJ2}, \rfr{sinpx} and \rfrs{elleAB}{accaAB} it straightforwardly turns out that
\begin{align}
\dot\Omega_\mathrm{LT}^\mathrm{B} & = \dot\Omega_\mathrm{LT}^\mathrm{A},\acap
\dot\Omega_{J_2}^\mathrm{B} & = -\dot\Omega_{J_2}^\mathrm{A},
\end{align}
which hold for an \textit{arbitrary} orientation of $\boldsymbol{\hat{k}}$ in space.

This proves that the orbital configurations by Ciufolini \citep{1986PhRvL..56..278C} and by Van Patten and Everitt \citep{1976CeMec..13..429V,1976PhRvL..36..629V}  are \textit{conceptually} equivalent, even \textit{regardless} of the \textit{inclination} of the orbital planes.
\subsection{The polar orbital configuration}
\begin{figure}
\centering
\includegraphics[width=\columnwidth]{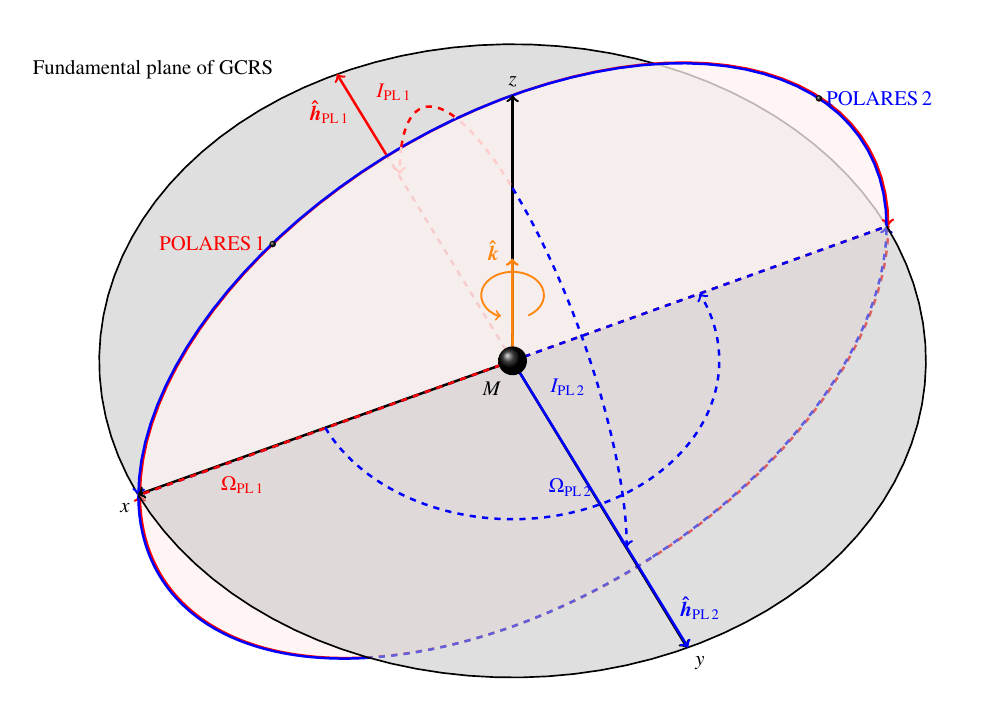}
\caption{Orbital configurations of POLARES 1 (in red) and POLARES 2 (in blue) at the epoch $T_1$ of the launch of LARES 2 with respect to the Geeocentric Celestial Reference System (GCRS) whose fundamental plane is shown shaded in grey. The Earth's spin axis $\boldsymbol{\hat{k}}$, in orange, is depicted according to \rfr{kT1}. The ratio of the semimajor axis $a$ of POLARES 2 to that of POLARES 1,  to which an arbitrary reference value has been assigned just for illustrative purpose, is as in Table \ref{Tab:1} from which the  eccentricities $e$ are retrieved as well. The inclinations $I$ and the longitudes of the ascending node $\Omega$ are given by \rfrs{polI}{polO} up to an offset of $\delta\gamma = \pm 1$ arcsecond for both orbital elements. The unit vectors $\boldsymbol{\hat{h}}$, given by \rfr{acca}, are directed towards the orbital angular momenta of the satellites. The sizes of the Earth and of the satellites' orbits are \textit{not} in scale. The view is from above the fundamental plane of GCRS.
}\label{fig_POLARES}
\end{figure}

\lb{polares}
For a generic orientation of $\boldsymbol{\hat{k}}$ in space, parameterized in terms of RA and decl. as per  \rfr{kappa}, the condition that the orbital plane contains the primary's spin axis is \textit{ideally} fulfilled if
\begin{align}
I \lb{polI} & = 90^\circ, \acap
\Omega \lb{polO} & = \alpha
\end{align}
\textit{exactly} hold.
Indeed, from \rfr{kappa} and \rfr{acca}, it turns out that \rfrs{polI}{polO} yield just
\eqi
\boldsymbol{\hat{h}}\boldsymbol\cdot\boldsymbol{\hat{k}} = 0.\lb{hk0}
\eqf
Figure \ref{fig_POLARES} depicts such a scenario, which may be branded as POLAr RElativity Satellites (POLARES).
For the sake of definiteness, the \textit{same} values as LAGEOS and LARES 2 were used for the semimajor axes and the eccentricities of the counter--revolving satellites POLARES 1 and 2, and the Earth's spin axis orientation was chosen as given by \rfr{kT1}. Furthermore, departures $\pm\delta \gamma$ with
\eqi
\delta\gamma = 1\,\mathrm{arcsecond} = 0.00027^\circ\lb{deltagamma}
\eqf
from the \textit{ideal} conditions of \rfrs{coI}{coO} and \rfrs{polI}{polO} were adopted.

\textcolor{black}{A peculiar advantage of a polar orbit configuration is that, \textit{in principle}, it allows to cancel out all the classical long-term rates of change of the node induced by the even zonal harmonics, as per \rfr{OJ2} and \rfr{hk0}; see also \citep{2024JGeod..98...77S}. }

\begin{figure}[ht!]
\centering
\begin{tabular}{c}
\includegraphics[width = 15 cm]{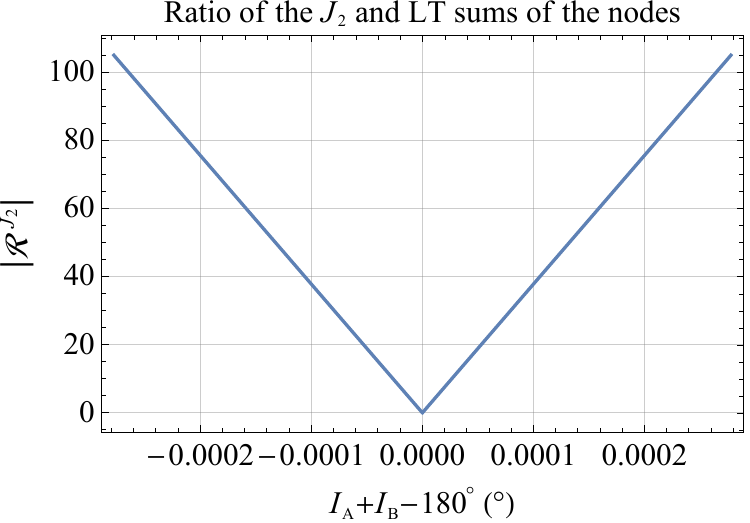}\\
\end{tabular}
\caption{
Plot of the absolute value of \rfr{RJ2}, calculated with \rfr{kT1}, \rfr{coO} and \rfrs{polI}{polO} in \rfrs{OLT}{OJ2}, as a function of the departure from the \textit{ideal} condition of \rfr{coI} within a range 2 arcseconds wide. The values for the semimajor axes and the eccentricities listed in Table \ref{Tab:1} were used.
}\label{Fig:2}
\end{figure}

\begin{figure}[ht!]
\centering
\begin{tabular}{c}
\includegraphics[width = 15 cm]{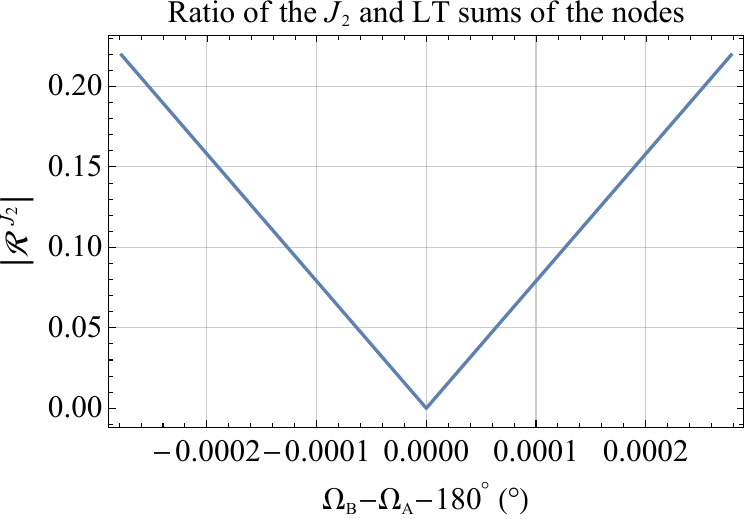}\\
\end{tabular}
\caption{Plot of the absolute value of \rfr{RJ2}, calculated with \rfr{kT1}, \rfr{coI} and \rfrs{polI}{polO} in \rfrs{OLT}{OJ2}, as a function of the departure from the \textit{ideal} condition of \rfr{coO} within a range 2 arcseconds wide. The values for the semimajor axes and the eccentricities listed in Table \ref{Tab:1} were used.
}\label{Fig:3}
\end{figure}

\begin{figure}[ht!]
\centering
\begin{tabular}{c}
\includegraphics[width = 15 cm]{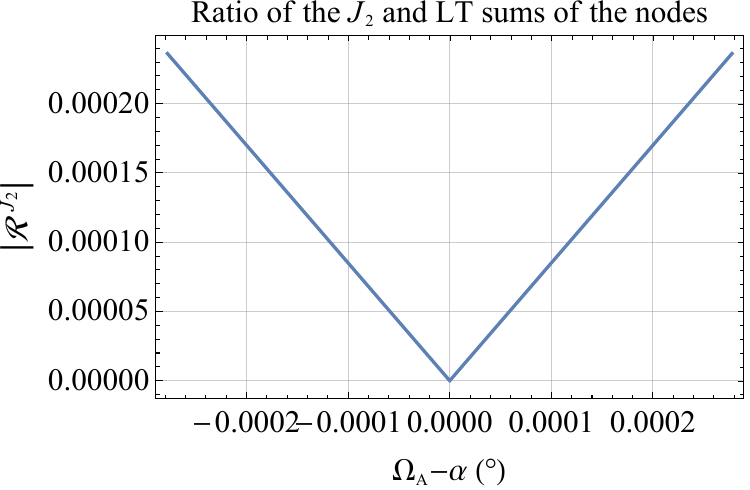}\\
\end{tabular}
\caption{Plot of the absolute value of \rfr{RJ2}, calculated with \rfr{kT1}, \rfrs{coI}{coO} and \rfr{polI} in \rfrs{OLT}{OJ2}, as a function of the departure from the \textit{ideal} condition of \rfr{polO} within a range 2 arcseconds wide. The values for the semimajor axes and the eccentricities listed in Table \ref{Tab:1} were used.
}\label{Fig:4}
\end{figure}

Figures \ref{Fig:2} to \ref{Fig:4} show that \rfr{RJ2}, calculated with \rfrs{OLT}{OJ2}, is \textit{less} sensitive to departures from the \textit{ideal} configuration established  by \rfrs{coI}{coO} and \rfrs{polI}{polO} than in the case of LAGEOS and LARES 2.
It turns out that, within the range given by \rfr{deltagamma}, $\mathcal{R}^{J_2}$ is mainly sensitive to the conditions of \rfr{coI} and \rfr{polI} on the orbital inclinations. Indeed, its maximum \textit{nominal} value can reach the level of about 100, which, however, is nearly \textit{600 times} \textit{smaller} than \rfr{morto} for LAGEOS and LARES 2. Instead, the conditions on the nodes of \rfr{coO} and \rfr{polO} can be somewhat relaxed, at least from the point of view of the reduction of the systematic bias due to $J_2$.

Thus, the earlier scenario envisaged by Van Patten and Everitt \citep{1976CeMec..13..429V,1976PhRvL..36..629V} is worth of being reconsidered. In view of the recent advances in accurate modeling the non--gravitaional perturbations and in the SLR technique, it may be implemented also with common geodetic satellites should the measurement of the LT effect be (\textit{one} of) its \textit{main} goal. Such a possibility would certainly deserve further investigations relying upon the past ones \citep{1976JAnSc..24..137S,1977JSpRo..14..474S,1978AcAau...5...77V,1982AcAau...9...55B}.
\section{Some comments on the criticisms raised by Ciufolini et al.}\lb{corbex}
Ciufolini \textit{et al.} \citep{Ciufoepjc24} raised certain criticisms to \citep{2023Univ....9..211I} which essentially boil down to the following.
\subsection{The issue of the error propagation}\lb{errpro}
Ciufolini \textit{et al.} \citep{Ciufoepjc24} \textit{repeatedly} accused Iorio \citep{2023Univ....9..211I} of ignoring even the most basic notions of error propagation in reaching his conclusions. Such an allegation would be based only (or mainly?) on the fact that Iorio \citep{2023Univ....9..211I} did not include (the \textit{static} part of) $J_2$ in the set of the parameters affected by errors to be propagated in $\mathcal{R}^{J_2}$.
Actually, it was just a matter of (sound) choice since, as correctly shown by Ciufolini \textit{et al.} \citep{Ciufoepjc24} themselves, the error of (the \textit{static} part of) $J_2$ impacts the overall uncertainty in \textit{the J2000.0  expression} of $\mathcal{R}^{J_2}$ adopted by Iorio \citep{2023Univ....9..211I} and Ciufolini \textit{et al.} \citep{Ciufoepjc24} to a negligible level.

On the contrary, it is precisely Ciufolini \textit{et al.} \citep{Ciufoepjc24} who seem to ignore how to properly propagate errors, judging by what they have done with the uncertainties on $a$ and $I$. Indeed, it is well known that if $f\ton{q_1,q_2,\ldots q_N}$ is an explicit function of $N$ parameters $q_i,\,i=1,2,\ldots N$ affected by experimental or observational errors $\sigma_{q_i},\,i=1,2,\ldots N$, an upper bound of the  total uncertainty in $f$ can be calculated as
\eqi
\sigma_f\leq \sum_{i=1}^N\left|\derp{f}{q_i}\right|\sigma_{q_i}.
\eqf
If $q_i,\,i=1,2,\ldots N$ are assumed to be mutually uncorrelated, then one can write
\eqi
\sigma_f = \sqrt{\sum_{i=1}^N\ton{\derp{f}{q_i}}^2\sigma^2_{q_i}}.
\eqf
It is precisely what Iorio \citep{2023Univ....9..211I} correctly did with $f\equiv\mathcal{R}^{J_2}$, assumed as function of the Newtonian constant of gravitation, of the Earth's standard gravitational parameter, equatorial radius, quadrupole mass moment and angular momentum, and of the semimajor axes, eccentricities and inclinations of LAGEOS and LARES 2 in assessing the impact of their uncertainties on $\mathcal{R}^{J_2}$ itself. Instead, Ciufolini \textit{et al.} \citep{Ciufoepjc24} incorrectly considered just the \textit{numerator} of $\mathcal{R}^{J_2}$, made of the sum of the classical node precessions, as a function of the aforementioned orbital elements, by keeping the \textit{denominator}, made of the sum of the LT node rates, \textit{fixed}. Instead, inexplicably, when it came to propagating the error on $J$, which enters the \textit{denominator} of $\mathcal{R}^{J_2}$, Ciufolini \textit{et al.} \citep{Ciufoepjc24} treated the latter as a \textit{variable} quantity dependent on $J$ itself by taking the derivative of the denominator \textit{only}. All this, together with the \textit{ad hoc} choice of the magnitudes of the errors in the orbital parameters, for the sole purpose of obtaining values more favorable to their preconceived assumptions. Suffice it to say that Ciufolini \textit{et al.} \citep{Ciufoepjc24}, after having criticized with dubious arguments the values adopted by Iorio \citep{2023Univ....9..211I} as representative of the experimental uncertainties in $a$, $e$ and $I$, in the end decided that $\sigma_{a_\mathrm{LR 2}}=0.1\,\mathrm{mm}$ by Iorio \citep{2023Univ....9..211I} was fine, while the figures by $\sigma_I$ by Iorio \citep{2023Univ....9..211I} himself were to be rejected; See the discussion about such a guess for $\sigma_a$ of the LAGEOS--type satellites in Section \ref{sema}.. Even from the strongly biased point of view of Ciufolini \textit{et al.} \citep{Ciufoepjc24} themselves, such an unjustified \virg{cherry--picking} attitude, as well as being contradictory and incorrect, is also counterproductive. Indeed, Iorio \citep{2023Univ....9..211I} showed that the bias on $\mathcal{R}^{J_2}$ due to $\sigma_e\simeq 10^{-5}$, tacitly accepted by Ciufolini \textit{et al.} \citep{Ciufoepjc24} since they did not criticize them, if \textit{correctly} calculated, amounts to $\simeq 120\%$ of the LT signature.
\subsection{The realistic assessment of the uncertainties in the Keplerian orbital parameters}\lb{bsciard}
About the correct evaluation of the uncertainties affecting the Keplerian orbital elements and their impact on the total error budget, their exceedingly small errors invoked by Ciufolini \textit{et al.} \citep{Ciufoepjc24} are likely the mere \textit{formal}, \textit{statistical} ones sorting out of the least--square estimation procedure of the satellites' data reduction. As such they are, by no means, representative of the \textit{physically realistic} uncertainties which have to be assessed from other pieces of information. Below, a pair of examples will be given.
\subsubsection{The semimajor axis}\lb{sema}
In the unperturbed Keplerian motion, the semimajor axis $a$ is calculated as
\eqi
a = \ton{\rp{2}{r} - \rp{v^2}{\upmu}}^{-1},\lb{sma}
\eqf
where $r$ and $v$ are the geocentric distance and speed of the satellite, respectively.
The relative error in $a$ due to the uncertainty in $\upmu$, averaged over one orbital period, turns out to be equal to the relative error in $\upmu$ itself \citep{2011GReGr..43.1697I} which, for the Earth, amounts to
\eqi
\rp{\sigma_\upmu}{\upmu}  = 1\times 10^{-9}.
\eqf
Such a figure is based on the online version of \cite[Chap. 1]{iers10} which quotes \citep{Ries07}.
Thus, the \textit{realistic} uncertainty in the semimajor axis of a LAGEOS--type satellite \textit{cannot} reasonably be smaller than
\eqi
\sigma_a\gtrsim \rp{\sigma_\upmu}{\upmu}a= 0.01\,\mathrm{m}=1\,\mathrm{cm},\lb{veroA}
\eqf
which is \textit{even} $1-2$ orders of magnitude \textit{larger} than the figures proposed in \citep{2023Univ....9..211I} and, for some reasons, accepted by Ciufolini \textit{et al.} \citep{Ciufoepjc24}.

By reassessing the impact of the uncertainties in $a$ on $\mathcal{R}^{J_2}$ with \rfr{veroA} yields
\eqi
\sigma_{\mathcal{R}^{J_2}}\leq \left|\derp{\mathcal{R}^{J_2}}{a_\mathrm{L}}\right|\sigma_{a_\mathrm{L}} + \left|\derp{\mathcal{R}^{J_2}}{a_\mathrm{LR\,2}}\right|\sigma_{a_\mathrm{LR\,2}}\simeq 1.1.
\eqf
\subsubsection{The inclination}\lb{inclis}
The fact that the an accuracy as high as  $0.01$ mas in the inclinations of LAGEOS and LARES 2 claimed by Ciufolini \textit{et al.} \citep{Ciufoepjc24} should be deemed as \textit{physically} unrealistic can be proven as follows.

The inclination $I$ the orbital plane of any satellite is reckoned from the principal plane of the ECI adopted, or, equivalently, from the latter's pole. Thus, the \textit{realistic} accuracy in $I$ \textit{cannot} certainly be \textit{better} than that of the reference directions themselves with respect to which it is defined.  The orientation of the ICRS, which coincides by default with that of GCRS, slightly differs from the MEME by a fixed offset expressed in terms of the three angles $\xi_0,\eta_0$ and $\mathrm{d}\alpha_0$; $\xi_0$ and $\eta_0$ are the celestial pole offsets at J2000 and $\mathrm{d}\alpha_0$ is the offset in right ascension of the J2000 mean equatorial frame with respect to the GCRS \citep{2006A&A...450..855C}. As far as the pole is concerned, according to \cite[Sect. 2.1.1]{iers10}, the discrepancy between different determinations of $\xi_0$ and $\eta_0$ is of the order of $\simeq 0.5$ mas and $\simeq 1.8$ mas, respectively; such figures can be reasonably assumed as representative of the uncertainty in the GCRS equator, at least due to the frame bias.
Furthermore, the direction of the ICRS pole is maintained fixed relative to the quasars within $\pm 20$ microarcseconds ($\mu$as) corresponding to $0.02$ mas \citep{1995A&A...303..604A,iers10,2024AJ....167..229M}. This shows that the \textit{physically meaningful} uncertainty in knowing the inclination $I$ of \textit{any} satellite cannot be better of
\eqi
\sigma_I\simeq 0.5-2\,\mathrm{mas}\lb{veroI}
\eqf

By reassessing the impact of the uncertainties in $I$ on $\mathcal{R}^{J_2}$ with \rfr{veroI} yields
\eqi
\sigma_{\mathcal{R}^{J_2}}\leq \left|\derp{\mathcal{R}^{J_2}}{I_\mathrm{L}}\right|\sigma_{I_\mathrm{L}} + \left|\derp{\mathcal{R}^{J_2}}{I_\mathrm{LR\,2}}\right|\sigma_{I_\mathrm{LR\,2}}\simeq 1.1.
\eqf
\subsection{The other criticisms by Ciufolini et al.}
All the rest of the criticisms by Ciufolini \textit{et al.} \citep{Ciufoepjc24}
don't make the case, completely bypassing the \textit{real} problem of the estimation of the LT effect in the data reduction raised in \cite[Sect.\,5]{2023Univ....9..211I}.

Suffice it to say that Ciufolini \textit{et al.} \citep{Ciufoepjc24}, in reply to \cite[Sect.\,5]{2023Univ....9..211I} who described in general terms the standard orbit determination procedure in satellite geodesy, astrodynamics and astronomy,  wrote that \virg{of course, in standard
space geodesy there is not such a thing as “simultaneously”
estimating the coefficients of the Earth gravitational field with “the propagation of electromagnetic waves”
and “the behaviour of measuring devices”}. Actually, in, e.g., \cite[p.\,1]{2004sod..book.....T}, one can just read: \virg{For the satellite orbit determination problem the minimal set of parameters will be the position and velocity vectors at some given epoch. In subsequent discussions, this minimal set will be expanded to include dynamic and measurement model parameters, which may be needed to improve the prediction accuracy.} Furthermore, in
\cite[p.\,3]{2004sod..book.....T} it is written: \virg{[\ldots] the orbit determination procedure may be used to obtain better estimates of the location of tracking stations, adjust the station clocks, calibrate radar biases, obtain a better estimate of geophysical constants, and so on.}

However, this is \textit{not} the important point, but rather the fact that, \textit{after almost} $\mathit{30}$ \textit{years} since the first attempts \citep{1996NCimA.109..575C},  neither Ciufolini and (past and present) coworkers nor anyone else have ever so far estimated a dedicated solve--for parameter of the LT effect in the least--square procedure of data reduction along with all the other parameters routinely estimated, whatever they may be, in satellite geodesy studies. This would be the \textit{only} significative breakdown in the  long history of LT tests with SLR. By repeating such estimations with different data sets and background reference models, and inspecting the covariance and correlation among \textit{all} the estimated parameters, \textit{including also a LT one} would be the \textit{only} correct and unambiguous way to proceed, as, on the other hand, already done by other teams in the Solar and Jovian scenarios \citep{2015IAUGA..2227771P,Pav2024,2011AGUFM.P41B1620F,2024ApJ...971..145D}.
Ciufolini \textit{et al.} \citep{Ciufoepjc24} did not offer \textit{any} answer at all to this important point. They limit themselves to cite just a few decades--old works whose authors have used times series of satellites' orbital elements to determine some non--gravitational physical effects, thus ignoring the \virg{evolution over the past four decades} \cite[p.\,1]{2004sod..book.....T} of the satellite orbit determination.
%
%
%
\section{Summary and conclusions}\lb{fine}
Recently, Ciufolini and coworkers firmly reaffirmed their belief about the possibility of successfully performing a $\simeq 0.2\%$ test of the post--Newtonian LT effect in the field of the Earth with the passive geodetic satellites LAGEOS and LARES 2 tracked with the Satellite Laser Ranging technique.

Should the terrestrial gravitomagnetic field  be explicitly \textit{modeled} and \textit{estimated} along with other parameters in dedicated satellites' data reductions, it would be possible to assess the overall systematic uncertainty in the \textit{standard way}, common to \textit{satellite geodesy}, \textit{astronomy} and \textit{astrodynamics},  by inspecting the \textit{correlations} among the \textit{estimated} LT parameter(s) and all the other determined ones contained in the covariance matrix of the fit.
It is worthwhile noticing that this \textit{is} just the way other teams of researchers recently performed their own LT tests with Mercury in the field of the Sun and with the probe Juno around Jupiter.
Such an approach should be repeated by \textit{varying} the \textit{data sets} of LAGEOS and LARES 2 themselves and the \textit{background reference models} adopted such as, e.g., different Earth's gravity field solutions produced by several institutions worldwide with data collected by dedicated spacecraft (GRACE, GOCE, GRACE-FO, other geodetic satellites) during different time spans. Given that, for unknown reasons, nearly \textit{30 years after} the first tests this has \textit{not} yet been done nor does it seem that it will be done in future tests, it is therefore more necessary than ever to resort to a \virg{\textit{offline}}, \textit{apriori} evaluation of the error budget based on \textit{analytical} methods and pieces of information collected in a variety of means.

The ratio of the sum of the Newtonian oblateness--driven node precessions of LAGEOS and LARES 2 to the sum of their LT counterparts fits well the scope. It can be viewed as a function of the orbital and physical parameters of the satellites and the Earth, respectively, all affected by observational uncertainties of various nature. \textit{In principle}, such a ratio would \textit{vanish} if the orbital sizes and shapes of both satellites  were \textit{identical} and the sum of their inclinations were \textit{precisely} $180^\circ$.

Recently, it was pointed out by the present author that the \textit{actual} mean values of the such orbital parameters of LAGEOS and LARES 2, averaged over a hundred days, do \textit{not} allow to meet this stringent goal, its resulting non--zero \textit{nominal} value being equal to almost $\mathit{5\,000}$. This unfortunate circumstance implies that the Earth's quadrupole moment sensibly impacts the proposed LT test also \textit{indirectly} through the errors in the physical and orbital parameters entering the aforementioned ratio.

Actually, previous estimates of such an important systematic bias were based  \textit{incorrectly} on formulas for the well known standard classical and relativistic node precessions which hold \textit{only} in an inertial reference system \textit{exactly} aligned with the Earth's spin axis. On the one hand, the former one is the Geocentric Celestial Reference System (GCRS), whose \textit{fixed} orientation in space \textit{nearly} coincides with the Earth's mean equator and equinox at the epoch \textit{J2000.0}. GCRS is just the inertial reference system used in  data analyses of Earth's satellites. On the other hand,  LARES 2 was launched about \textit{22 years after} that epoch. Then, \textit{general} formulas for the node precessions of interest, valid for an \textit{arbitrary} orientation of the \textit{precessing} Earth's spin axis, should be used.
By repeating the above analytical calculation for the epoch of the launch of  LARES 2 by \textit{consistently propagating} the errors in \textit{all} the parameters entering the ratio of the summed classical to relativistic precessions shows that its \textit{nominal} value is \textit{even larger} than its J2000.0 counterpart, amounting now to about $\mathit{59\,000}$. Thus, the \textit{indirect} impact of the inaccurately known physical and orbital parameters entering it is \textit{even larger} than in the J2000.0 case recently investigated in the literature. The uncertainties in parameters such as $J_2$ itself and the Earth's angular momentum $J$ which did not have a remarkable impact in the J2000.0 case, now also play a non--negligible role--to the percent order--in pushing the \textit{realistically} achievable accuracy away from the final $0.2\%$ accuracy goal sought for the LT test with LAGEOS and LARES 2. Among them, the impact of the secular rate of change of the Earth's quadrupole mass moment should deserve a careful evaluation.

The \textit{exceedingly small} errors in the orbital parameters of LAGEOS and LARES 2 claimed by Ciufolini and  coworkers are \textit{unlikely} representative of any \textit{realistic}, \textit{physically meaningful} uncertainty in them, being just the \textit{mere statistical} errors of the fitting procedure performed in the data reduction. The present--day relative uncertainty in the Earth's standard gravitational parameter entering the calculation of the semimajor axis $a$ does \textit{not} allow to \textit{realistically} know it with an accuracy \textit{better} than about \textit{1 centimetre}. The satellite's orbital inclination $I$, being, by definition, reckoned from the GCRS pole axis, can only be known with an accuracy \textit{necessarily} limited by that of the orientation of GCRS itself, being the latter of the order of $\simeq \mathit{0.5-2}$ \textit{milliarcseconds}.

It has been shown that, as far as the \textit{orbital configuration} is concerned, the earlier proposal by Van Patten and Everitt of two counter--orbiting drag--free satellites in \textit{identical}  polar orbits is \textit{conceptually equivalent} to the LAGEOS--LARES 2 one in the sense that \textit{both} imply that, \textit{in principle}, the sum of the LT node precessions add up while the classical ones due to $J_2$ cancel out, irrespectively of their altitudes and eccentricities provided that they are \textit{ideally} equal for both the hypothesized spacecraft. Notably, this feature does \textit{not} necessarily hold \textit{only} for polar orbits.

Furthermore, it turns out that a pair of counter--revolving satellites in \textit{nearly} equal orbits both passing  \textit{almost} exactly through the Earth's poles would represent a scenario \textit{less sensitive} to the impact of the unavoidable departures from the \textit{idealized} one than LAGEOS and LARES 2. Indeed, by assuming for the sake of definiteness the \textit{same} orbital sizes and shapes of the latter ones, the \textit{nominal} ratio of the summed classical to relativistic node precessions would reach a maximum of \textit{just 100} for deviations of the inclinations from their \textit{ideal} values of up to \textit{1 arcseconds}.

The Van Patten--Everitt proposal, revamped and rebranded POLAr RElativity Satellites (POLARES), may be implemented even with passive geodetic satellites tracked with  the SLR technique only.
\section*{Data availability}
No new data were generated or analysed in support of this research.
\section*{Conflict of interest statement}
I declare no conflicts of interest.
\bibliography{Megabib}{}

\begin{thebibliography}{78}
\providecommand{\natexlab}[1]{#1}
\providecommand{\url}[1]{\texttt{#1}}
\expandafter\ifx\csname urlstyle\endcsname\relax
  \providecommand{\doi}[1]{doi: #1}\else
  \providecommand{\doi}{doi: \begingroup \urlstyle{rm}\Url}\fi

\bibitem[{Arias} et~al.(1995){Arias}, {Charlot}, {Feissel}, and
  {Lestrade}]{1995A&A...303..604A}
E.~F. {Arias}, P.~{Charlot}, M.~{Feissel}, and J.~F. {Lestrade}.
\newblock {The extragalactic reference system of the International Earth
  Rotation Service, ICRS.}
\newblock \emph{Astron. Astrophys.}, 303:\penalty0 604--608, 1995.

\bibitem[{Bogorodskii}(1959)]{Bogo59}
A.~F. {Bogorodskii}.
\newblock Relativistic effects in the motion of an artificial earth satellite.
\newblock \emph{Soviet Astron.}, 3:\penalty0 857--862, 1959.

\bibitem[{Bolton} et~al.(2017){Bolton}, {Lunine}, {Stevenson},
  et~al.]{2017SSRv..213....5B}
S.~J. {Bolton}, J.~{Lunine}, D.~{Stevenson}, et~al.
\newblock {The Juno Mission}.
\newblock \emph{Space Sci. Rev.}, 213:\penalty0 5--37, 2017.
\newblock \doi{10.1007/s11214-017-0429-6}.

\bibitem[{Breakwell} et~al.(1982){Breakwell}, {Everitt}, {Schaechter}, and {Van
  Patten}]{1982AcAau...9...55B}
J.~V. {Breakwell}, C.~W.~F. {Everitt}, D.~B. {Schaechter}, and R.~A. {Van
  Patten}.
\newblock {Geodesy information in a modified relativity mission with two
  counter-orbiting polar satellites}.
\newblock \emph{Acta Astronaut.}, 9:\penalty0 55--56, Jan 1982.
\newblock \doi{10.1016/0094-5765(82)90033-9}.

\bibitem[{Capderou}(2005)]{2005som..book.....C}
M.~{Capderou}.
\newblock \emph{{Satellites: Orbits and missions}}.
\newblock Springer, 2005.

\bibitem[{Capitaine} and {Wallace}(2006)]{2006A&A...450..855C}
N.~{Capitaine} and P.~T. {Wallace}.
\newblock {High precision methods for locating the celestial intermediate pole
  and origin}.
\newblock \emph{Astron. Astrophys.}, 450:\penalty0 855--872, 2006.
\newblock \doi{10.1051/0004-6361:20054550}.

\bibitem[{Ciufolini}(1986)]{1986PhRvL..56..278C}
I.~{Ciufolini}.
\newblock {Measurement of the Lense-Thirring drag on high-altitude,
  laser-ranged artificial satellites}.
\newblock \emph{Phys. Rev. Lett.}, 56:\penalty0 278--281, 1986.
\newblock \doi{10.1103/PhysRevLett.56.278}.

\bibitem[{Ciufolini}(1996)]{1996NCimA.109.1709C}
I.~{Ciufolini}.
\newblock {On a new method to measure the gravitomagnetic field using two
  orbiting satellites.}
\newblock \emph{Nuovo Cim. A}, 109A:\penalty0 1709--1720, 1996.
\newblock \doi{10.1007/BF02773551}.

\bibitem[{Ciufolini} et~al.(1996){Ciufolini}, {Lucchesi}, {Vespe}, and
  {Mandiello}]{1996NCimA.109..575C}
I.~{Ciufolini}, D.~M. {Lucchesi}, F.~{Vespe}, and A.~{Mandiello}.
\newblock {Measurement of dragging of inertial frames and gravitomagnetic field
  using laser--ranged satellites.}
\newblock \emph{Nuovo Cim. A}, 109A:\penalty0 575--590, 1996.
\newblock \doi{10.1007/BF02731140}.

\bibitem[{Ciufolini} et~al.(1997){Ciufolini}, {Chieppa}, {Lucchesi},
  et~al.]{1997CQGra..14.2701C}
I.~{Ciufolini}, F.~{Chieppa}, D.~M. {Lucchesi}, et~al.
\newblock {Test of Lense - Thirring orbital shift due to spin}.
\newblock \emph{Class. Quantum Gravit.}, 14:\penalty0 2701--2726, 1997.
\newblock \doi{10.1088/0264-9381/14/10/003}.

\bibitem[{Ciufolini} et~al.(1998){Ciufolini}, {Pavlis}, {Chieppa},
  et~al.]{1998Sci...279.2100C}
I.~{Ciufolini}, E.~C. {Pavlis}, F.~{Chieppa}, et~al.
\newblock {Test of General Relativity and Measurement of the Lense-Thirring
  Effect with Two Earth Satellites}.
\newblock \emph{Science}, 279:\penalty0 2100, 1998.
\newblock \doi{10.1126/science.279.5359.2100}.

\bibitem[{Ciufolini} et~al.(2023){Ciufolini}, {Paolozzi}, {Pavlis},
  et~al.]{2023EPJC...83...87C}
I.~{Ciufolini}, A.~{Paolozzi}, E.~C. {Pavlis}, et~al.
\newblock {The LARES 2 satellite, general relativity and fundamental physics}.
\newblock \emph{Eur. Phys. J. C}, 83:\penalty0 87, 2023.
\newblock \doi{10.1140/epjc/s10052-023-11230-6}.

\bibitem[{Ciufolini} et~al.(2024){Ciufolini}, {Paris}, {Pavlis},
  et~al.]{Ciufoepjc24}
I.~{Ciufolini}, C.~{Paris}, E.~C. {Pavlis}, et~al.
\newblock {On the high accuracy to test dragging of inertial frames with the
  LARES 2 space experiment}.
\newblock \emph{Eur. Phys. J. C}, 84:\penalty0 998, 2024.
\newblock \doi{10.1140/epjc/s10052-024-13301-8}.

\bibitem[{Cohen} and {Smith}(1985)]{1985JGR....90.9217C}
S.~C. {Cohen} and D.~E. {Smith}.
\newblock {LAGEOS Scientific results: Introduction}.
\newblock \emph{J. Geophys. Res.}, 90:\penalty0 9217--9220, 1985.
\newblock \doi{10.1029/JB090iB11p09217}.

\bibitem[{Coulot} et~al.(2011){Coulot}, {Deleflie}, {Bonnefond}, et~al.]{SLR11}
D.~{Coulot}, F.~{Deleflie}, P.~{Bonnefond}, et~al.
\newblock Satellite laser ranging.
\newblock In H.~K {Gupta}, editor, \emph{Encyclopedia of Solid Earth
  Geophysics}, Encyclopedia of Earth Sciences Series, pages 1049--1055.
  Springer, 2011.
\newblock \doi{10.1007/978-90-481-8702-7\textunderscore 98}.

\bibitem[{Cugusi} and {Proverbio}(1978)]{1978A&A....69..321C}
L.~{Cugusi} and E.~{Proverbio}.
\newblock {Relativistic Effects on the Motion of Earth's Artificial
  Satellites}.
\newblock \emph{Astron. Astrophys.}, 69:\penalty0 321--325, 1978.

\bibitem[{Durante} et~al.(2024){Durante}, {Cappuccio}, {di Stefano},
  et~al.]{2024ApJ...971..145D}
D.~{Durante}, P.~{Cappuccio}, I.~{di Stefano}, et~al.
\newblock {Testing General Relativity with Juno at Jupiter}.
\newblock \emph{Astrophys. J.}, 971:\penalty0 145, 2024.
\newblock \doi{10.3847/1538-4357/ad5ff5}.

\bibitem[{Dymnikova}(1986)]{1986SvPhU..29..215D}
I.~G. {Dymnikova}.
\newblock {REVIEWS OF TOPICAL PROBLEMS: Motion of particles and photons in the
  gravitational field of a rotating body (In memory of Vladimir Afanas'evich
  Ruban)}.
\newblock \emph{Sov. Phys. Usp.}, 29:\penalty0 215--237, 1986.
\newblock \doi{10.1070/PU1986v029n03ABEH003178}.

\bibitem[{Everitt}(1974)]{Varenna74}
C.~W.~F. {Everitt}.
\newblock {The Gyroscope experiment - I: General description and analysis of
  gyroscope performance}.
\newblock In B.~{Bertotti}, editor, \emph{Proceedings of the International
  School of Physics \virg{Enrico Fermi}. Course LVI. Experimental Gravitation},
  pages 331--360. Academic Press, 1974.

\bibitem[{Everitt} et~al.(2001){Everitt}, {Buchman}, {Debra},
  et~al.]{2001LNP...562...52E}
C.~W.~F. {Everitt}, S.~{Buchman}, D.~B. {Debra}, et~al.
\newblock {Gravity Probe B: Countdown to Launch}.
\newblock In C.~{L{\"a}mmerzahl}, C.~W.~F. {Everitt}, and F.~W. {Hehl},
  editors, \emph{Gyros, Clocks, Interferometers ...: Testing Relativistic
  Gravity in Space}, volume 562 of \emph{Lecture Notes in Physics}, pages
  52--82. Springer, 2001.
\newblock \doi{10.1007/3-540-40988-2\textunderscore 4}.

\bibitem[{Everitt} et~al.(2011){Everitt}, {Debra}, {Parkinson},
  et~al.]{2011PhRvL.106v1101E}
C.~W.~F. {Everitt}, D.~B. {Debra}, B.~W. {Parkinson}, et~al.
\newblock {Gravity Probe B: Final Results of a Space Experiment to Test General
  Relativity}.
\newblock \emph{Phys. Rev. Lett.}, 106:\penalty0 221101, 2011.
\newblock \doi{10.1103/PhysRevLett.106.221101}.

\bibitem[{Everitt} et~al.(2015){Everitt}, {Muhlfelder}, {Debra},
  et~al.]{2015CQGra..32v4001E}
C.~W.~F. {Everitt}, B.~{Muhlfelder}, D.~B. {Debra}, et~al.
\newblock {The Gravity Probe B test of general relativity}.
\newblock \emph{Class. Quantum Gravit.}, 32:\penalty0 224001, 2015.
\newblock \doi{10.1088/0264-9381/32/22/224001}.

\bibitem[{Finocchiaro} et~al.(2011){Finocchiaro}, {Iess}, {Folkner}, and
  {Asmar}]{2011AGUFM.P41B1620F}
S.~{Finocchiaro}, L.~{Iess}, W.~M. {Folkner}, and S.~{Asmar}.
\newblock {The Determination of Jupiter's Angular Momentum from the
  Lense-Thirring Precession of the Juno Spacecraft}.
\newblock In \emph{AGU Fall Meeting Abstracts}, volume 2011, pages P41B--1620,
  2011.

\bibitem[{Ginzburg}(1957)]{Ginz57a}
V.~L. {Ginzburg}.
\newblock The use of artificial earth satellites for verifying the general
  theory of relativity.
\newblock \emph{Usp. Fiz. Nauk}, 63:\penalty0 119--122, 1957.

\bibitem[{Ginzburg}(1959)]{Ginz59}
V.~L. {Ginzburg}.
\newblock {Artificial Satellites and the Theory of Relativity}.
\newblock \emph{Sci. Am.}, 200:\penalty0 149--160, 1959.
\newblock \doi{10.1038/scientificamerican0559-149}.

\bibitem[{Heiskanen} and {Moritz}(1967)]{1967phge.book.....H}
W.~A. {Heiskanen} and H.~{Moritz}.
\newblock \emph{{Physical geodesy}}.
\newblock Freeman, 1967.

\bibitem[{Ibba} et~al.(1989){Ibba}, {Rum}, {Varesio}, and
  {Bussolino}]{1989AcAau..19..521I}
R.~{Ibba}, G.~{Rum}, F.~{Varesio}, and L.~{Bussolino}.
\newblock {IRIS--LAGEOS 2 mission}.
\newblock \emph{Acta Astronaut.}, 19:\penalty0 521--527, 1989.
\newblock \doi{10.1016/0094-5765(89)90119-7}.

\bibitem[{Iorio}(2003)]{2003CeMDA..86..277I}
L.~{Iorio}.
\newblock {The Impact of the Static Part of the Earth's Gravity Field on Some
  Tests of General Relativity with Satellite Laser Ranging}.
\newblock \emph{Celest. Mech. Dyn. Astr.}, 86:\penalty0 277--294, 2003.
\newblock \doi{10.1023/A:1024223200686}.

\bibitem[{Iorio}(2011)]{2011GReGr..43.1697I}
L.~{Iorio}.
\newblock {How accurate is the cancelation of the first even zonal harmonic of
  the geopotential in the present and future LAGEOS-based Lense-Thirring
  tests?}
\newblock \emph{Gen. Relativ. Gravit.}, 43:\penalty0 1697--1706, 2011.
\newblock \doi{10.1007/s10714-011-1151-4}.

\bibitem[{Iorio}(2023)]{2023Univ....9..211I}
L.~{Iorio}.
\newblock {Limitations in Testing the Lense{\textendash}Thirring Effect with
  LAGEOS and the Newly Launched Geodetic Satellite LARES 2}.
\newblock \emph{Universe}, 9:\penalty0 211, 2023.
\newblock \doi{10.3390/universe9050211}.

\bibitem[{Iorio}(2024)]{2024gpno.book.....I}
L.~{Iorio}.
\newblock \emph{{General Post-Newtonian Orbital Effects From Earth's Satellites
  to the Galactic Center}}.
\newblock Cambridge University Press, 2024.
\newblock \doi{10.1017/9781009562911}.

\bibitem[{Iorio} et~al.(2002){Iorio}, {Lucchesi}, and
  {Ciufolini}]{2002CQGra..19.4311I}
L.~{Iorio}, D.~M. {Lucchesi}, and I.~{Ciufolini}.
\newblock {The LARES mission revisited: an alternative scenario}.
\newblock \emph{Class. Quantum Gravit.}, 19:\penalty0 4311--4325, 2002.
\newblock \doi{10.1088/0264-9381/19/16/307}.

\bibitem[{Iorio} et~al.(2011){Iorio}, {Lichtenegger}, {Ruggiero}, and
  {Corda}]{2011Ap&SS.331..351I}
L.~{Iorio}, H.~I.~M. {Lichtenegger}, M.~L. {Ruggiero}, and C.~{Corda}.
\newblock {Phenomenology of the Lense--Thirring effect in the solar system}.
\newblock \emph{Astrophys. Space Sci.}, 331:\penalty0 351--395, 2011.
\newblock \doi{10.1007/s10509-010-0489-5}.

\bibitem[{Kopeikin} et~al.(2011){Kopeikin}, {Efroimsky}, and
  {Kaplan}]{2011rcms.book.....K}
S.~M. {Kopeikin}, M.~{Efroimsky}, and G.~{Kaplan}.
\newblock \emph{{Relativistic Celestial Mechanics of the Solar System}}.
\newblock Wiley, 2011.
\newblock \doi{10.1002/9783527634569}.

\bibitem[{Lense} and {Thirring}(1918)]{1918PhyZ...19..156L}
J.~{Lense} and H.~{Thirring}.
\newblock {{\"U}ber den Einflu{\ss} der Eigenrotation der Zentralk{\"o}rper auf
  die Bewegung der Planeten und Monde nach der Einsteinschen
  Gravitationstheorie}.
\newblock \emph{Phys. Z}, 19:\penalty0 156--163, 1918.

\bibitem[{Lucchesi}(2001)]{2001P&SS...49..447L}
D.~M. {Lucchesi}.
\newblock {Reassessment of the error modelling of non--gravitational
  perturbations on LAGEOS II and their impact in the Lense--Thirring
  determination. Part I}.
\newblock \emph{Planet. Space Sci.}, 49:\penalty0 447--463, 2001.
\newblock \doi{10.1016/S0032-0633(00)00168-9}.

\bibitem[{Lucchesi}(2002)]{2002P&SS...50.1067L}
D.~M. {Lucchesi}.
\newblock {Reassessment of the error modelling of non--gravitational
  perturbations on LAGEOS II and their impact in the Lense--Thirring
  derivation--Part II}.
\newblock \emph{Planet. Space Sci.}, 50:\penalty0 1067--1100, 2002.
\newblock \doi{10.1016/S0032-0633(02)00052-1}.

\bibitem[{Lucchesi}(2004)]{2004CeMDA..88..269L}
D.~M. {Lucchesi}.
\newblock {LAGEOS Satellites Germanium Cube--Corner-Retroreflectors and the
  Asymmetric Reflectivity Effect}.
\newblock \emph{Celest. Mech. Dyn. Astr.}, 88:\penalty0 269--291, 2004.
\newblock \doi{10.1023/B:CELE.0000017171.78328.f1}.

\bibitem[{Lucchesi} et~al.(2004){Lucchesi}, {Ciufolini}, {Andr{\'e}s},
  et~al.]{2004P&SS...52..699L}
D.~M. {Lucchesi}, I.~{Ciufolini}, J.~I. {Andr{\'e}s}, et~al.
\newblock {LAGEOS II perigee rate and eccentricity vector excitations residuals
  and the Yarkovsky--Schach effect}.
\newblock \emph{Planet. Space Sci.}, 52:\penalty0 699--710, 2004.
\newblock \doi{10.1016/j.pss.2004.01.007}.

\bibitem[{Malkin}(2024)]{2024AJ....167..229M}
Z.~{Malkin}.
\newblock {How Well is the International Celestial Reference System Maintained
  in Official IAU Implementations?}
\newblock \emph{Astron J.}, 167\penalty0 (5):\penalty0 229, 2024.
\newblock \doi{10.3847/1538-3881/ad35bf}.

\bibitem[{Mashhoon}(2001)]{2001rfg..conf..121M}
B.~{Mashhoon}.
\newblock {Gravitoelectromagnetism}.
\newblock In J.~F. {Pascual-S{\'a}nchez}, L.~{Flor{\'\i}a}, A.~{San Miguel},
  and F.~{Vicente}, editors, \emph{Reference Frames and Gravitomagnetism}.
  World Scientific, 2001.

\bibitem[{Mashhoon}(2007)]{Mash07}
B.~{Mashhoon}.
\newblock {Gravitoelectromagnetism: A Brief Review}.
\newblock In L.~{Iorio}, editor, \emph{The Measurement of Gravitomagnetism: A
  Challenging Enterprise}, pages 29--39. Nova Science, 2007.

\bibitem[{Mashhoon} et~al.(1984){Mashhoon}, {Hehl}, and
  {Theiss}]{1984GReGr..16..711M}
B.~{Mashhoon}, F.~W. {Hehl}, and D.~S. {Theiss}.
\newblock {On the gravitational effects of rotating masses: The Thirring--Lense
  papers.}
\newblock \emph{Gen. Relativ. Gravit.}, 16:\penalty0 711--750, 1984.
\newblock \doi{10.1007/BF00762913}.

\bibitem[{Milani} et~al.(1987){Milani}, {Nobili}, and
  {Farinella}]{Nobilibook87}
A.~{Milani}, A.M. {Nobili}, and P.~{Farinella}.
\newblock \emph{{Non--gravitational perturbations and satellite geodesy}}.
\newblock Adam Hilger, 1987.

\bibitem[{Montenbruck} and {Gill}(2000)]{2000Monte}
O.~{Montenbruck} and E.~{Gill}.
\newblock \emph{{Satellite Orbits}}.
\newblock Spinger-Verlag, Berlin Heidelberg, 2000.
\newblock \doi{10.1007/978-3-642-58351-3}.

\bibitem[{Paolozzi} et~al.(2019){Paolozzi}, {Sindoni}, {Felli},
  et~al.]{2019JGeod..93.2437P}
A.~{Paolozzi}, G.~{Sindoni}, F.~{Felli}, et~al.
\newblock {Studies on the materials of LARES 2 satellite}.
\newblock \emph{J. Geod.}, 93:\penalty0 2437--2446, 2019.
\newblock \doi{10.1007/s00190-019-01316-z}.

\bibitem[{Park} et~al.(2015){Park}, {Folkner}, and
  {Konopliv}]{2015IAUGA..2227771P}
R.~S. {Park}, W.~M. {Folkner}, and A.~S. {Konopliv}.
\newblock {Estimation of Solar Angular Momentum from Lense--Thirring Precession
  of Mercury}.
\newblock \emph{IAU General Assembly}, 22:\penalty0 2227771, 2015.

\bibitem[{Pavlov} and {Dolgakov}(2024{\natexlab{a}})]{Pav2024}
D.~{Pavlov} and I.~{Dolgakov}.
\newblock {General relativity tests by the dynamics of the Solar system}.
\newblock In C.~{Bizouard}, A.~{Fienga}, and F.~{Paganelli}, editors,
  \emph{Proceedings of the Journ\'{e}es 2023 \virg{Syst\`{e}mes de
  rf\'{e}r\'{e}nce spatio-temporels}}, pages 156--160. Observatoire de la
  C\^{o}te d'Azur, 2024{\natexlab{a}}.

\bibitem[{Pavlov} and {Dolgakov}(2024{\natexlab{b}})]{RussiLT}
D.~{Pavlov} and I.~{Dolgakov}.
\newblock {Studying the Properties of Spacetime with an Improved Dynamical
  Model of the Inner Solar System}.
\newblock \emph{Universe}, 10:\penalty0 413, 2024{\natexlab{b}}.
\newblock \doi{10.3390/universe10110413}.

\bibitem[{Pearlman} et~al.(2019){Pearlman}, {Arnold}, {Davis},
  et~al.]{2019JGeod..93.2181P}
M.~{Pearlman}, D.~{Arnold}, M.~{Davis}, et~al.
\newblock {Laser geodetic satellites: a high-accuracy scientific tool}.
\newblock \emph{J. Geod.}, 93:\penalty0 2181--2194, 2019.
\newblock \doi{10.1007/s00190-019-01228-y}.

\bibitem[{Pearlman} et~al.(2002){Pearlman}, {Degnan}, and
  {Bosworth}]{2002AdSpR..30..135P}
M.~R. {Pearlman}, J.~J. {Degnan}, and J.~M. {Bosworth}.
\newblock {The International Laser Ranging Service}.
\newblock \emph{Adv. Space Res.}, 30:\penalty0 135--143, 2002.
\newblock \doi{10.1016/S0273-1177(02)00277-6}.

\bibitem[{Petit} and {Luzum}(2010)]{iers10}
G.~{Petit} and B.~{Luzum}, editors.
\newblock \emph{IERS Conventions (2010)}, volume~36 of \emph{IERS Technical
  Note}.
\newblock Verlag des Bundesamts f\"{u}r Kartographie und Geod\"{a}sie,
  Frankfurt am Main, 2010.

\bibitem[{Pfister}(2007)]{2007GReGr..39.1735P}
H.~{Pfister}.
\newblock {On the history of the so--called Lense--Thirring effect}.
\newblock \emph{Gen. Relativ. Gravit.}, 39:\penalty0 1735--1748, 2007.
\newblock \doi{10.1007/s10714-007-0521-4}.

\bibitem[{Pfister}(2008)]{2008mgm..conf.2456P}
H.~{Pfister}.
\newblock {The History of the So--Called Lense--Thirring Effect}.
\newblock In H.~{Kleinert}, R.~T. {Jantzen}, and R.~{Ruffini}, editors,
  \emph{The Eleventh Marcel Grossmann Meeting On Recent Developments in
  Theoretical and Experimental General Relativity, Gravitation and Relativistic
  Field Theories}, pages 2456--2458. World Scientific, 2008.
\newblock \doi{10.1142/9789812834300\textunderscore 0433}.

\bibitem[{Pfister}(2014)]{Pfister2014}
H.~{Pfister}.
\newblock {Gravitomagnetism: From Einstein's 1912 Paper to the Satellites
  LAGEOS and Gravity Probe B}.
\newblock In J.~{Bi\v{c}\'{a}k} and T.~{Ledvinka}, editors, \emph{Relativity
  and Gravitation}, volume 157 of \emph{Springer Proceedings in Physics,},
  pages 191--197. Springer, 2014.
\newblock \doi{10.1007/978-3-319-06761-2\textunderscore 24}.

\bibitem[{Poisson} and {Will}(2014)]{2014grav.book.....P}
E.~{Poisson} and C.~M. {Will}.
\newblock \emph{{Gravity. Newtonian, Post--Newtonian, Relativistic}}.
\newblock Cambridge University Press, 2014.
\newblock \doi{10.1017/CBO9781139507486}.

\bibitem[{Pugh}(1959)]{Pugh59}
G.~E. {Pugh}.
\newblock {Proposal for a Satellite Test of the Coriolis Prediction of General
  Relativity}.
\newblock Research memorandum, Weapons Systems Evaluation Group, The Pentagon,
  Washington D.C., 1959.

\bibitem[{Renzetti}(2013)]{2013CEJPh..11..531R}
G.~{Renzetti}.
\newblock {History of the attempts to measure orbital frame--dragging with
  artificial satellites}.
\newblock \emph{Centr. Eur. J. Phys.}, 11:\penalty0 531--544, 2013.
\newblock \doi{10.2478/s11534-013-0189-1}.

\bibitem[{Ries}(2007)]{Ries07}
J.~C. {Ries}.
\newblock {Satellite laser ranging and the terrestrial reference frame;
  Principal sources of uncertainty in the determination of the scale}.
\newblock In \emph{Geophysical Research Abstracts}, volume~9 of \emph{EGU
  General Assembly Conference Abstracts}, page 10809, 2007.

\bibitem[{Ries} et~al.(2016){Ries}, {Bettadpur}, {Eanes}, et~al.]{Riesetal16}
J.~C. {Ries}, S.~V. {Bettadpur}, R.~J. {Eanes}, et~al.
\newblock {The Development and Evaluation of the Global Gravity Model GGM05}.
\newblock CSR Technical Report CSR-16-02, Center for Space Research, 2016.

\bibitem[{Rindler}(2001)]{2001rsgc.book.....R}
W.~{Rindler}.
\newblock \emph{{Relativity: special, general, and cosmological}}.
\newblock Oxford University Press, 2001.

\bibitem[{Ruggiero} and {Tartaglia}(2002)]{2002NCimB.117..743R}
M.~L. {Ruggiero} and A.~{Tartaglia}.
\newblock {Gravitomagnetic effects}.
\newblock \emph{Nuovo Cim. B}, 117:\penalty0 743, 2002.

\bibitem[{Schaechter} et~al.(1976){Schaechter}, {Breakwell}, {Van Patten}, and
  {Everitt}]{1976JAnSc..24..137S}
D.~{Schaechter}, J.~V. {Breakwell}, R.~A. {Van Patten}, and F.~W. {Everitt}.
\newblock {Collision avoidance for two counter-orbiting satellites.}
\newblock \emph{J. Astronaut. Sci.}, 24:\penalty0 137--146, Jun 1976.

\bibitem[{Schaechter} et~al.(1977){Schaechter}, {Breakwell}, {Van Patten}, and
  {Everitt}]{1977JSpRo..14..474S}
D.~{Schaechter}, J.~V. {Breakwell}, R.~A. {Van Patten}, and C.~W.~F. {Everitt}.
\newblock {Covariance analysis for a relativity mission.}
\newblock \emph{J. Spacecr. Rockets}, 14:\penalty0 474--478, 1977.
\newblock \doi{10.2514/3.57226}.

\bibitem[{Sch{\"a}fer}(2004)]{2004GReGr..36.2223S}
G.~{Sch{\"a}fer}.
\newblock {Gravitomagnetic Effects}.
\newblock \emph{Gen. Relativ. Gravit.}, 36:\penalty0 2223--2235, 2004.
\newblock \doi{10.1023/B:GERG.0000046180.97877.32}.

\bibitem[{Sch{\"a}fer}(2009)]{2009SSRv..148...37S}
G.~{Sch{\"a}fer}.
\newblock {Gravitomagnetism in Physics and Astrophysics}.
\newblock \emph{Space Sci. Rev.}, 148:\penalty0 37--52, 2009.
\newblock \doi{10.1007/s11214-009-9537-2}.

\bibitem[{Schiff}(1960)]{Schiff60}
L.~{Schiff}.
\newblock {Possible new experimental test of general relativity theory}.
\newblock \emph{Phys. Rev. Lett.}, 4:\penalty0 215--217, 1960.
\newblock \doi{10.1103/PhysRevLett.4.215}.

\bibitem[{Shapiro}(1990)]{1990grg..conf..313S}
I.~I. {Shapiro}.
\newblock {Solar system tests of general relativity: recent results and present
  plans}.
\newblock In N.~{Ashby}, D.~F. {Bartlett}, and W.~{Wyss}, editors,
  \emph{General Relativity and Gravitation, 1989}, pages 313--330. Cambridge
  University Press, 1990.
\newblock \doi{10.1017/CBO9780511564178.025}.

\bibitem[{So{\'s}nica}(2024)]{2024JGeod..98...77S}
K.~{So{\'s}nica}.
\newblock {Orbit design for a future geodetic satellite and gravity field
  recovery}.
\newblock \emph{J. Geod.}, 98:\penalty0 77, 2024.
\newblock \doi{10.1007/s00190-024-01884-9}.

\bibitem[{Souchay} and {Capitaine}(2013)]{PrecNut}
J.~{Souchay} and N.~{Capitaine}.
\newblock {Precession and Nutation of the Earth}.
\newblock In J.~{Souchay}, S.~{Mathis}, and T.~{Tokieda}, editors, \emph{Tides
  in Astronomy and Astrophysics}, volume 861 of \emph{Lecture Notes in
  Physics}, pages 115--166. Springer, 2013.
\newblock \doi{10.1007/978-3-642-32961-6\textunderscore 4}.

\bibitem[{Tapley} et~al.(2004){Tapley}, {Schutz}, and
  {Born}]{2004sod..book.....T}
B.~D. {Tapley}, B.~E. {Schutz}, and G.~H. {Born}.
\newblock \emph{{Statistical Orbit Determination}}.
\newblock Elsevier, 2004.
\newblock \doi{10.1016/B978-0-12-683630-1.X5019-X}.

\bibitem[{Thorne}(1986)]{1986hmac.book..103T}
K.~S. {Thorne}.
\newblock {Black Holes: The Membrane Viewpoint}.
\newblock In S.~L. {Shapiro}, S.~A. {Teukolsky}, and E.~E. {Salpeter}, editors,
  \emph{Highlights of Modern Astrophysics: Concepts and Controversies}, pages
  103--161. Wiley, 1986.

\bibitem[{Thorne}(1988)]{1988nznf.conf..573T}
K.~S. {Thorne}.
\newblock {Gravitomagnetism, jets in quasars, and the Stanford Gyroscope
  Experiment.}
\newblock In J.~D. {Fairbank}, Jr. {Deaver}, B.~S., C.~W.~F. {Everitt}, and
  P.~F. {Michelson}, editors, \emph{Near Zero: New Frontiers of Physics}, pages
  573--586. Freeman, 1988.

\bibitem[{Thorne} et~al.(1986){Thorne}, {MacDonald}, and {Price}]{Thorne86}
K.~S. {Thorne}, D.~A. {MacDonald}, and R.~H. {Price}, editors.
\newblock \emph{{Black Holes: The Membrane Paradigm}}.
\newblock Yale University Press, 1986.

\bibitem[{Torge}(2001)]{Torge01}
W.~{Torge}.
\newblock \emph{{Geodesy. 3rd Edition}}.
\newblock de Gruyter, 2001.

\bibitem[{Van Patten} and {Everitt}(1976{\natexlab{a}})]{1976CeMec..13..429V}
R.~A. {Van Patten} and C.~W.~F. {Everitt}.
\newblock {A Possible Experiment with Two Counter-Orbiting Drag-Free Satellites
  to Obtain a New Test of Einstein's General Theory of Relativity and Improved
  Measurements in Geodesy}.
\newblock \emph{Celest. Mech. Dyn. Astr.}, 13:\penalty0 429--447,
  1976{\natexlab{a}}.
\newblock \doi{10.1007/BF01229096}.

\bibitem[{Van Patten} and {Everitt}(1976{\natexlab{b}})]{1976PhRvL..36..629V}
R.~A. {Van Patten} and C.~W.~F. {Everitt}.
\newblock {Possible experiment with two counter-orbiting drag-free satellites
  to obtain a new test of Einstein's general theory of relativity and improved
  measurements in geodesy}.
\newblock \emph{Phys. Rev. Lett.}, 36:\penalty0 629--632, 1976{\natexlab{b}}.
\newblock \doi{10.1103/PhysRevLett.36.629}.

\bibitem[{Van Patten} et~al.(1978){Van Patten}, {Breakwell}, {Schaechter}, and
  {Everitt}]{1978AcAau...5...77V}
R.~A. {Van Patten}, J.~V. {Breakwell}, D.~{Schaechter}, and C.~W.~F. {Everitt}.
\newblock {Error analysis of a relativity test with counterorbiting
  satellites.}
\newblock \emph{Acta Astronaut.}, 5:\penalty0 77--86, 1978.
\newblock \doi{10.1016/0094-5765(78)90036-X}.

\end{thebibliography}
\end{document}